\documentclass[aps,pre,reprint,
  superscriptaddress,
  longbibliography,
  floatfix
]{revtex4-2}

\usepackage{amsmath,amssymb,bm}
\usepackage{graphicx}
\usepackage[hidelinks]{hyperref}
\usepackage{here}
\usepackage{color}
\usepackage{tabularx}
\usepackage{booktabs}
\usepackage{soul}

\newcommand{\tr}{\operatorname{tr}}
\newcommand{\dev}{\operatorname{dev}}

\begin{document}

\title{A deviatoric-stress closure for constitutive modeling of viscoelastic dynamics}
\author{Souta Miyamoto}
\email{s.miyamoto@cheme.kyoto-u.ac.jp}
\affiliation{Department of Chemical Science and Engineering, Kyoto University,
	Kyoto 615-8510, Japan}
\affiliation{School of Chemical Science and Technology, Kyoto University, Kyoto 615-8510,
Japan}
\author{Shotaro Moro}
\affiliation{School of Chemical Science and Technology, Kyoto University, Kyoto 615-8510,
Japan}
\author{Takeshi Sato}
\affiliation{Advanced Manufacturing Technology Institute, Kanazawa University, Kanazawa 920-1192, Japan}
\author{Shota Kato}
\affiliation{Department of Informatics, Kyoto University, Kyoto 606-8501, Japan}
\author{Katsuaki Tanabe}
\affiliation{Department of Chemical Science and Engineering, Kyoto University,
	Kyoto 615-8510, Japan}
\affiliation{School of Chemical Science and Technology, Kyoto University, Kyoto 615-8510,
Japan}
\affiliation{Kyoto MPI Inc., Kyoto 606-8501, Japan}
\author{John J. Molina}
\affiliation{Department of Chemical Science and Engineering, Kyoto University,
	Kyoto 615-8510, Japan}
\affiliation{School of Chemical Science and Technology, Kyoto University, Kyoto 615-8510,
Japan}
\author{Takashi Taniguchi}
\affiliation{Department of Chemical Science and Engineering, Kyoto University,
	Kyoto 615-8510, Japan}
\affiliation{School of Chemical Science and Technology, Kyoto University, Kyoto 615-8510,
Japan}
\date{\today}
\begin{abstract}
Standard rheological measurements yield only selected stress components; thus, inferring tensorial constitutive equations from experimentally accessible observables is complicated. 
We propose a constitutive formulation written in terms of a deviatoric stress tensor, whose trace is zero, rather than the extra stress tensor. 
From rheometric data including shear stress, first and second normal stress differences under shear, and elongational stress under uniaxial elongation, we can construct a deviatoric stress state without the indeterminate isotropic stress.
The deviatoric-stress dynamics is represented by a closure inferred through symbolic regression, constrained to satisfy material objectivity and a given linear Maxwell response.
To demonstrate the proposed formulation, two closures inferred from stress responses of the Giesekus and Larson models successfully captured untrained transient-flow responses under planar elongation and mixed shear/uniaxial elongations at deformation rates around an inverse relaxation time. 
Steady rheological functions of the closures agreed with the original models in the linear-response regime and over a deformation-rate range connected to the training data, whereas deviations and divergent responses appeared at larger deformation rates outside the training regime. 
These results demonstrate that the proposed deviatoric-stress formulation provides a practical route for constitutive modeling of observable linear and nonlinear viscoelastic dynamics, while clarifying its range of validity under strong deformation.
\end{abstract}

\maketitle

\section{Introduction}
\label{sec:intro}

Constitutive models describing observed stress responses have long been investigated in rheology~\cite{Doi1988-yf,Larson1998-ia}.
Conventional constitutive equations are usually formulated as tensorial stress equations, based on phenomenological or molecular pictures.
Various constitutive models have been proposed, such as the Oldroyd framework~\cite{Oldroyd1958-jh}, the Giesekus model~\cite{Giesekus1982-ov}, the Phan-Thien--Tanner (PTT) model~\cite{Thien1977-gs}, the Larson model~\cite{Larson1984-hw}, and the finite extensible nonlinear elastic model with Peterlin's closure (FENE-P)~\cite{Peterlin1966-xc}.
These descriptions basically cover single-mode relaxation dynamics, and their linear and nonlinear stress responses under different deformation modes have been summarized~\cite{Hyun2011-xf,Bharadwaj2015-ov,Saengow2015-np,Saengow2018-hx,Saengow2019-uw,Song2020-ln}. 
Nevertheless, selecting an appropriate model from rheometric data is difficult because some combinations of models with specific parameters can produce similar responses under limited deformation protocols~\cite{Saadat2022-lk,Dabiri2023-st,John2024-bu}.
Thus, model selection, parameter calibration, and physical interpretation rely on expert knowledge~\cite{Saadat2023-pg}.

Recent progress in data-driven methods~\cite{Brunton2019-gp,Vinuesa2022-bm,Mangal2024-gi} has provided another route for constructing such constitutive equations, while incorporating basic physical constraints.
Various data-driven approaches for inferring viscoelastic models have been proposed for different settings, including approaches based on measured rheological responses~\cite{Mahmoudabadbozchelou2022-er,Mahmoudabadbozchelou2024-zi}, identification of tensorial stress dynamics~\cite{Seryo2020-ib,Miyamoto2023-ot,Zhao2021-kl,Jin2023-ki,Lennon2023-tt,Shanbhag2024-po,Sato2025-yx,Sato2025-eh,Rodrigues2025-ke,Mangal2025-fo,Sato2026-fk}, and learning based on latent physical descriptors such as conformation tensors~\cite{Fang2024-vx,Dong2026-ic,Park2026-zw}.
These studies demonstrate the potential of data-driven approaches as ``digital twins'' for rheometers and as predictors of complex fluid flows.

Nevertheless, constructing a tensorial constitutive equation from experimentally accessible stress responses of a material remains challenging.
Constitutive equations are conventionally written in terms of the extra stress tensor in three-dimensional space (which excludes hydrostatic pressure contributions), though experiments under shear or elongational deformation usually provide only a few combinations of stress components, such as the shear stress, normal stress differences, and elongational stress. 
Considerable effort has been devoted to providing high-precision rheological data under various flow types. 
For example, a cone-partitioned plate setup can be used to measure the second normal stress difference~\cite{Schweizer2013-om,Costanzo2018-sc,Costanzo2024-cb}.
Extensional rheological responses have also been measured such as by capillary-thinning methods~\cite{McKinley2000-ow}, and by stagnation-point microfluidic devices under planar, uniaxial, and biaxial elongations~\cite{Haward2012-px,Haward2023-bd,Haward2023-lg}.
However, these advanced measurements only provide limited stress information rather than the complete second rank stress tensor with its independent six components.

The difficulty is that extra-stress formulations contain an isotropic trace degree of freedom, that is undetermined by standard rheological measurements.
In these formulations, the isotropic degree of freedom contributes to nonlinear stress responses, for example, explicitly in the PTT and FENE-P models.
We note that Shanbhag and Erlebacher~\cite{Shanbhag2024-po} pointed out the non-uniqueness of constitutive models inferred from partial measurements, using oscillatory shear as an example.
This non-uniqueness problem makes it difficult to select an appropriate model from experimental data.
Thus, we here develop a formulation based on the traceless stress tensor, namely the deviatoric stress tensor, as the state variable.

In this study, we propose a deviatoric-stress-based formulation as a reduced-state closure, and develop a data-driven inference method from standard rheological measurements. 
The proposed formulation builds on the tensorial formulation proposed by Lennon et al.~\cite{Lennon2023-tt} and is consistent with material objectivity and a linear Maxwell response.
The reduced-state tensorial formulation can be regarded as an effective description at the level of measured stress responses, without requiring estimation of the unobserved isotropic stress component.
To demonstrate the proposed formulation, we examine whether the observable stress dynamics generated by known constitutive equations can be reproduced.
In the subsequent section, we explain the theoretical formulation.

\begin{figure}
    \centering
    \includegraphics[width=0.95\linewidth]{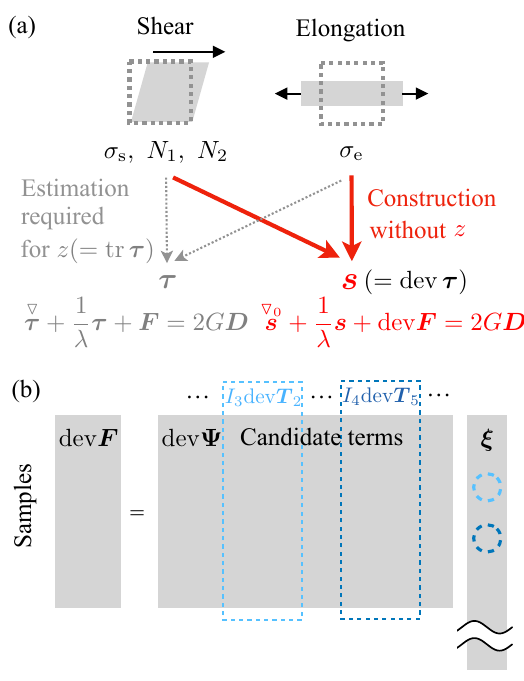}
    \caption{
    Conceptual illustration of the proposed method based on the deviatoric-stress tensor ($\bm s$), in comparison with the conventional formulation based on the extra-stress tensor ($\bm\tau$).
    (a) Maxwell-type formulations based on the extra stress tensor (left) and the deviatoric stress tensor (right).
    The upper panels show two rheological deformations accessible in experiments.
    Shear measurements provide the shear stress ($\sigma_{\rm s}$) and the first and second normal stress differences ($N_1$ and $N_2$), whereas uniaxial elongation provides the elongational stress ($\sigma_{\rm e}$).
    The extra stress tensor $\bm{\tau}$ cannot be reconstructed from these observables without estimating the additional isotropic component $z=\tr \bm{\tau}$.
    By contrast, under the symmetry of the corresponding deformation, the deviatoric stress tensor
    $\bm{s}=\dev \bm{\tau}$ is directly reconstructed from the measured stress components, and its dynamics can be inferred. 
    The definitions of the equations for $\bm\tau$ and $\bm s$, and those of the symbols are given in Eqs.~\eqref{eq:model-nondim} and \eqref{eq:reduced-closure}.
    The overset $\triangledown_0$ of $\bm s$ is defined as Eq.~\eqref{eq:traceless-derivative}.
    (b) Regression problem of the additional deviatoric term $\dev \bm{F}$ in the deviatoric-stress-based constitutive equation.
    The library matrix $\bm{\Psi}$ is constructed from products of the scalar invariants $I_i$ and the deviatoric part of the basis tensors $\dev \bm{T}_i$ obtained from the representation theorem (Eq.~\eqref{eq:objective-rep}), and $\bm{\xi}$ denotes the coefficient vector, as defined in Eq.~\eqref{eq:F-sindy}.
    }
    \label{fig:schematic-illustration}
\end{figure}

\section{Model formulation}
\label{sec:model}
In this section we review the extra-stress formulation in Sec.~\ref{sec:totalstress-formula} and then derive its reduction to a deviatoric-stress closure in Sec.~\ref{sec:reduced-formula}.
A latent-state formulation for partially observed extra-stress dynamics is given in Appendix~\ref{sec:latent-estimation} as an alternative way, but it is not used in the main analysis because of its numerical difficulty.

\subsection{Extra-stress formulation}
\label{sec:totalstress-formula}

We consider the viscoelastic constitutive model~\cite{Lennon2023-tt,Shanbhag2024-po,Rodrigues2025-ke}, in which the extra-stress tensor $\bm{\tau}$ evolves under the velocity gradient $\bm{\kappa}$. 
The strain-rate tensor is defined as $\bm{D}=(\bm{\kappa}+\bm{\kappa}^{\mathsf{T}})/2$, where $\bm{\kappa}=(\bm{\nabla}\bm{u})^{\mathsf{T}}$ and $\bm{u}$ is the velocity field.
To account for contributions not captured by the upper-convected Maxwell (UCM) model, an additional term $\bm{F}$ is introduced as
\begin{equation}
\overset{\triangledown}{\bm{\tau}}+\frac{1}{\lambda}\bm{\tau}+\bm{F}
=2G\bm{D},
\label{eq:UDE}
\end{equation}
where $G$ is the modulus and $\lambda$ is the relaxation time, and $\overset{\triangledown}{\bm{\tau}}$ is the upper-convected derivative of $\bm{\tau}$, defined as $\overset{\triangledown}{\bm{\tau}}=\dot{\bm \tau}- \bm\tau\cdot\bm{\kappa}^{\mathsf{T}}-\bm{\kappa}\cdot\bm{\tau}$.
The advection term $\bm{u}\cdot\nabla\bm{\tau}$ is absent for the homogeneous deformations considered here.
We nondimensionalize stress and time by $G$ and $\lambda$, respectively, and use the same symbols for the dimensionless variables, yielding
\begin{equation}
\overset{\triangledown}{\bm{\tau}}
+\bm{\tau}
+\bm{F}
=2\bm{D}.
\label{eq:model-nondim}    
\end{equation}

Material objectivity requires $\bm F$ to be an isotropic tensor function~\cite{Spencer2004-es} of $\bm{\tau}$ and $\bm{D}$, such that a simultaneous rotation of the inputs induces the same rotation of the output.
Using the Cayley--Hamilton theorem, we can write the tensor function $\bm F$ as~\cite{Lennon2023-tt}
\begin{equation}
\bm{F}(\bm{\tau},\bm{D})
=\sum_{i=1}^{9}g_i(I_1(\bm \tau,\bm D),\ldots,I_9(\bm \tau,\bm D))\,\bm{T}_i(\bm \tau,\bm D),
\label{eq:objective-rep}    
\end{equation}
where $g_i$ is a scalar function of the invariants $\{I_i(\bm \tau,\bm D)\}_{i=1,\cdots,9}$,
and $\{\bm{T}_i(\bm \tau,\bm D)\}_{i=1,\cdots,9}$ form a tensor basis. 
The basis tensors  $\bm{T}_i$ and scalar invariants $I_i$ are summarized in Table~\ref{tab:objective-basis}.
Here, incompressibility, $\mathrm{tr}\,\bm D=0$, is assumed.
We note that this choice is not unique (i.e., $\bm \tau\cdot\bm D \cdot \bm\tau$ can be used instead of ${\bm T}_7$).

\begin{table}[H]
\caption{Tensor basis $\{\bm{T}_i\}_{i=1,\cdots,9}$ and scalar invariants $\{I_i\}_{i=1,\cdots,9}$ representing isotropic $\bm F(\bm\tau,\bm D)$. \textbf{I} is the unit tensor.}
\label{tab:objective-basis}
\centering
\begin{tabular}{lll}
\toprule
$i\quad$&Tensor basis $\bm{T}_i(\bm \tau,\bm D)$ & Invariants $I_i(\bm \tau,\bm D)$ \\
\midrule
1&$\textbf{I}$ & $\tr\bm{\tau}$ \\
2&$\bm{\tau}$ & $\tr(\bm{\tau}\cdot\bm{\tau})$ \\
3&$\bm{D}$ & $\tr(\bm{D}\cdot\bm{D})$ \\
4&$\bm{\tau}\cdot\bm{\tau}$ & $\tr(\bm{\tau}\cdot\bm{\tau}\cdot\bm{\tau})$ \\
5&$\bm{D}\cdot\bm{D}$ & $\tr(\bm{D}\cdot\bm{D}\cdot\bm{D})$ \\
6&$\bm{\tau}\cdot\bm{D}+\bm{D}\cdot\bm{\tau}$ & $\tr(\bm{\tau}\cdot\bm{D})$ \\
7&$\bm{\tau}\cdot\bm{\tau}\cdot\bm{D}+\bm{D}\cdot\bm{\tau}\cdot\bm{\tau}$ & $\tr(\bm{\tau}\cdot\bm{\tau}\cdot\bm{D})$ \\
8&$\bm{\tau}\cdot\bm{D}\cdot\bm{D}+\bm{D}\cdot\bm{D}\cdot\bm{\tau}$ & $\tr(\bm{\tau}\cdot\bm{D}\cdot\bm{D})$ \\
9&$\bm{\tau}\cdot\bm{\tau}\cdot\bm{D}\cdot\bm{D}+\bm{D}\cdot\bm{D}\cdot\bm{\tau}\cdot\bm{\tau}$ & $\tr(\bm{\tau}\cdot\bm{\tau}\cdot\bm{D}\cdot\bm{D})$ \\
\bottomrule
\end{tabular}
\end{table}

The representation in Eq.~\eqref{eq:objective-rep} includes several standard constitutive equations as special cases.
For example, the Giesekus~\cite{Giesekus1982-ov}, PTT~\cite{Thien1977-gs}, Larson~\cite{Larson1984-hw}, FENE-P~\cite{Peterlin1966-xc}, and Oldroyd 8-constant~\cite{Oldroyd1958-jh,Giacomin2022-sy} models are written as
\begin{align}
    \text{Giesekus:}\,
    \bm F(\bm\tau,\bm D)&=\xi_\mathrm{G}\bm T_4, \label{eq:F_Giesekus}\\
    \text{Larson:}\,
    \bm F(\bm\tau,\bm D)&=\frac{2}{3}\xi_\mathrm{L} I_6(\bm T_1+\bm T_2), \label{eq:F_Larson}\\
    \text{PTT:}\,
    \bm F(\bm\tau,\bm D)&=[\exp(\varepsilon_\mathrm{P} I_1)-1]\bm T_2
    +\xi_\mathrm{P}\bm T_6, \label{eq:F_PTT}\\
    \text{FENE-P:}\,
    \bm F(\bm\tau,\bm D)&=\frac{f(I_1)I_1-2I_6}{3b^2}(\bm T_1+\bm T_2)\nonumber\\
    & \qquad -(1-f(I_1))\bm T_2, \label{eq:F_FENE-P}\\
    \text{Oldroyd-8:}\,
    \bm F(\bm\tau,\bm D)&=c_{31}I_3\bm T_1+c_{61}I_6\bm T_1+c_3\bm T_3\nonumber\\
    &\qquad +c_{13}I_1\bm T_3+c_5\bm T_5+c_6\bm T_6. \label{eq:F_Oldroyd-8}
\end{align}
Here, $\xi_{\mathrm{G}}$, $\xi_{\mathrm{L}}$, $\varepsilon_{\mathrm{P}}$, and $\xi_{\mathrm{P}}$ are model parameters, and $c_{31}$, $c_{61}$, $c_3$, $c_{13}$, $c_5$, and $c_6$ are coefficients for each term for the Oldroyd 8-constant model.
The parameter $b$ appearing in Eq.~\eqref{eq:F_FENE-P} is the maximum stretching ratio in the FENE-P model, and $f(I_1)=(3b^2+3+I_1)/(3b^2)$.
These terms represent polymeric stress contributions and are decoupled from purely viscous stress contributions.

However, inferring the contributions of these extra-stress-based terms from standard rheological measurements remains nontrivial.
One major reason is that not all nonzero components of $\bm{\tau}$ are observable in terms of standard rheological experiments.
For example, under simple shear, the three measurable quantities are the shear stress, the first normal stress difference, and the second normal stress difference.
These quantities do not uniquely determine the four nonzero components of $\bm{\tau}$, namely the three diagonal components and one shear component.
The remaining latent degree of freedom is $\tr\bm{\tau}$, although this trace mode appears in several extra-stress-based terms listed above.
This observation motivates the partially observed extra-stress formulation described in Appendix~\ref{sec:latent-estimation}.
However, simultaneously inferring nonlinear constitutive terms and estimating the latent trace mode is numerically difficult.
Thus, we reduce the model state by eliminating $\tr\bm{\tau}$, as described in the following section.

\subsection{Deviatoric-stress closure}
\label{sec:reduced-formula}

We introduce a reduced constitutive model written in terms of the deviatoric stress tensor $\bm{s}$.
Decomposing $\bm\tau$ into its trace and deviatoric parts, $z$ and $\bm s$, respectively, we can obtain the following expressions:
\begin{align}
    \bm \tau = \bm s+\frac{1}{3}z\textbf{I},
\end{align}
where
\begin{align}
    z=\tr\bm\tau,\quad \bm s=\dev\bm\tau=\bm\tau-\frac{1}{3}(\tr\bm\tau)\textbf{I}.    
\end{align}
We use the notation $\dev\bm X$ for the traceless part of a second rank tensor $\bm X$ as $\dev \bm X=\bm X - (\tr\bm X)\textbf{I}/3$.
By substituting this decomposition into Eq.~\eqref{eq:model-nondim}, the time evolutions of $z$ and $\bm{s}$ are respectively given by
\begin{align}
 &\dot{z} + z + \tr{\bm F}(\bm s+\frac{1}{3}z\textbf{I},\bm D)= 2\bm D:\bm s, \\
 &\overset{\triangledown_0}{\bm s} +\bm s
 +\dev{\bm F}(\bm s+\frac{1}{3}z\textbf{I},\bm D) = 2\left(1+\frac{z}{3}\right)\bm D, \label{eq:deviatoric-dynamics}
\end{align}
where
\begin{equation}
    \overset{\triangledown_0}{\bm s}=\overset{\triangledown}{\bm s}+\frac{2}{3}(\bm s:\bm D)\textbf{I}.
    \label{eq:traceless-derivative}
\end{equation}
Equation~\eqref{eq:deviatoric-dynamics} shows that the dynamics of $\bm{s}$ is still coupled to $z$.
Since $z$ is not independently determined from standard rheological measurements, we define a minimal observable closure by taking $\bm{s}$ as the state variable.
We introduce the following ansatz by removing the explicit $z$-dependence from the deviatoric dynamics:
\begin{equation}
\overset{\triangledown_0}{\bm s} +\bm s+\dev{\bm F}(\bm s,\bm D) = 2\bm D.
\label{eq:reduced-closure}
\end{equation}
This ansatz gives a minimal closed constitutive model for $\bm{s}$.
It can be viewed as being obtained by replacing $\bm{\tau}$ with $\bm{s}$ in Eq.~\eqref{eq:model-nondim} and considering the term $(2/3)(\bm{s}:\bm{D})\textbf{I}$ to preserve the traceless condition, $\mathrm{tr}\,\dot{\bm{s}}=0$.
The resulting closure satisfies material objectivity and recovers the linear viscoelastic response in the small-deformation limit.

Note that this ansatz should be interpreted as a closure for observable deviatoric-stress dynamics, not as an exact projection of the original extra-stress equation. 
In the full decomposition, the trace variable $z$ affects the deviatoric dynamics through $-2z\bm{D}/3$ and through the $z$-dependence of $\bm{F}$, but these contributions are not directly determined from standard rheological observables. 
Equation~\eqref{eq:reduced-closure} avoids imposing an underdetermined $z$-dynamics and defines a minimal closed model of $\bm s$. 
As shown in Appendix~\ref{sec:trace-elimination-UCM}, even the extra-stress UCM model, for which $\bm F=\bm 0$, yields a memory contribution after exact elimination of $z$.

Following this reduction procedure, we consider two baseline projected closures, the projected Giesekus and Larson models:
\begin{align}
    \text{Projected Giesekus:}\,
    &\dev \bm F(\bm s,\bm D)=\xi_\mathrm{G}\dev(\bm s\cdot\bm s),
    \label{eq:projected-giesekus}   \\
    \text{Projected Larson:}\, 
    &\dev \bm F(\bm s,\bm D)=\frac{2}{3}\xi_\mathrm{L} \tr(\bm s\cdot\bm D)\bm s.\label{eq:projected-larson}
\end{align}
The steady shear and uniaxial elongational responses of these projected closures are discussed in Sec.~\ref{sec:analysis-on-reduced-formula}.

We then consider data-driven inference of reduced closures from stress responses, generated by the original Giesekus and Larson models.
By a data-driven method, the eliminated trace contributions are translated into $\dev\bm{F}(\bm{s},\bm{D})$ in the reduced closure.
The numerical formulation and setup for this inference problem are described in Sec.~\ref{sec:numerical-methods}.

In the following numerical studies, the state $\bm{s}$ is constructed from stress variables accessible in rheological measurements.
Under shear with $\dot{\gamma}=\kappa_{xy}$, we assume that the observable variables are the shear stress $\sigma_\mathrm{s}=s_{xy}$, the first normal stress difference $N_1=s_{xx}-s_{yy}$, and the second normal stress difference $N_2=s_{yy}-s_{zz}$.
The deviatoric stress tensor is then reconstructed as
\begin{align}
\bm s=
\begin{bmatrix}
\displaystyle{\frac{2N_1+N_2}{3}}&\sigma_\mathrm{s}&0 \\
\sigma_\mathrm{s}&\displaystyle{\frac{-N_1+N_2}{3}}& 0\\
0&0&\displaystyle{-\frac{N_1+2N_2}{3}}
\end{bmatrix}.
\label{eq:dev-stress-from-shear}
\end{align}
Under uniaxial elongation with $\kappa_{xx}=\dot{\varepsilon}$ and $\kappa_{yy}=\kappa_{zz}=-\dot{\varepsilon}/2$, the uniaxial elongational stress $\sigma_\mathrm{e}=s_{xx}-(s_{yy}+s_{zz})/2$ gives
\begin{align}
\bm s=
\begin{bmatrix}
\displaystyle{\frac{2}{3}\sigma_\mathrm{e}}&0&0 \\
0&\displaystyle{-\frac{1}{3}\sigma_\mathrm{e}}& 0\\
0&0&\displaystyle{-\frac{1}{3}\sigma_\mathrm{e}}
\end{bmatrix}.
\label{eq:dev-stress-from-elongation}
\end{align}

\section{Numerical methods}
\label{sec:numerical-methods}

The symbolic-regression formulation for inferring reduced closure dynamics is described in Sec.~\ref{sec:symbolic-regression}.
The numerical setup, including the deformation protocols, candidate library, and model-selection procedure, is described in Sec.~\ref{sec:numerical-setup}.

\subsection{Symbolic-regression formulation}
\label{sec:symbolic-regression}

Following symbolic-regression approaches for constitutive modeling~\cite{Shanbhag2024-po,Sato2025-eh,Sato2025-yx} based on the nonlinear dynamics identification framework~\cite{Brunton2016-ie}, we formulate a linear regression problem for inferring $\bm{F}$ while considering the nonlinearity.
To cast $\bm{F}$ into a linear regression form, we expand each scalar coefficient $g_i$ in a prescribed library of invariant-based candidate functions $\{\phi_{ij}\}_{j=1,\cdots,M_i}$, where $M_i$ is the number of candidates associated with $\bm{T}_i$. 
Substitution of the expansion into Eq.~\eqref{eq:objective-rep} gives
\begin{align}
\bm{F}(\bm{s},\bm{D})
&=\sum_{i=1}^{9}g_i(I_1(\bm s,\bm D),\ldots,I_9(\bm s,\bm D))\,\bm{T}_i(\bm s,\bm D)
\\
&=
\sum_{i=1}^{9}\sum_{j=1}^{M_i}
\xi_{ij}\,\phi_{ij}(I_1,\dots,I_9)\,\bm{T}_i \\
&=
\sum_{m=1}^{M}\xi_m\,\bm{\Psi}_m,
\label{eq:F-sindy}
\end{align}
where $M=\sum_{i=1}^9 M_i$, $\bm{\Psi}_m=\phi_{ij}(I_1,\dots,I_9)\bm{T}_i$, and $m$ the flattened single index from the pair $(i,j)$, i.e., $(i,j) \rightarrow m=j+\sum_{k=1}^{i-1}M_k$ (where the sum vanishes for $i=1$). 
$\xi_{ij}$ and $\xi_m$ denote the same coefficient.
To avoid numerical problems arising from higher order terms, a simple choice of first order polynomials is used in this study.
For example, when one chooses the $M_2=3$ candidate functions as ($1$, $I_2$, $I_3$) for $i=2$, $\phi_{21}=1$, $\phi_{22}=I_2$, and $\phi_{23}=I_3$ are the invariant-based functions to be multiplied with $\bm{T}_2$.

The same procedure can be applied to the extra-stress formulation in Eq.~\eqref{eq:model-nondim}, but the main analysis below uses the reduced dynamics shown in Eq.~\eqref{eq:reduced-closure}.
Given a set of $N$ samples, $\{\bm{\kappa}^{(n)}, \bm{s}^{(n)}, \dot{\bm{s}}^{(n)} \}_{n=1,\ldots,N}$, we construct the corresponding regression data ($\bm\kappa$ is used in the upper-convected time derivative).
For each sample, $\dev\bm F^{(n)}$ is evaluated as the residual of Eq.~\eqref{eq:reduced-closure}, and the candidate term $\bm{\Psi}^{(n)}_m$ is evaluated from $\bm{\kappa}^{(n)}$ and $\bm{s}^{(n)}$.

Given these data, the linear regression problem is formulated as follows. 
The coefficient vector $\bm{\xi} = [\xi_1, \ldots, \xi_M]^\mathsf{T}$ is determined by minimizing the objective 
\begin{align}
J(\bm{\xi})
&=
\frac{1}{N}\sum_{n=1}^N
\left\|
\bm{f}^{(n)} - \bm{\theta}^{(n)} \bm{\xi}
\right\|_2^2
+ \mathcal{R}(\bm{\xi}),
\end{align}
where
\begin{align}
\mathrm{vec}_{\mathrm{s}}(\bm{A})
&=
[A_{xx}\ A_{yy}\ A_{zz}\ \sqrt{2}A_{xy}\ \sqrt{2}A_{yz}\ \sqrt{2}A_{zx}]^\mathsf{T}, \label{eq:vec-sym}\\
\bm{f}^{(n)} 
&=
\mathrm{vec}_{\mathrm{s}}(\dev\bm{F}^{(n)}), \\
\bm{\theta}^{(n)}
&=
\bigl[
\mathrm{vec}_{\mathrm{s}}(\dev\bm{\Psi}_1^{(n)})\,
\ldots\,
\mathrm{vec}_{\mathrm{s}}(\dev\bm{\Psi}_M^{(n)})
\bigr].
\end{align}
Here, $\|\cdot\|_2$ corresponds to the $\ell_2$ norm of the input vector.
$\mathrm{vec}_{\mathrm{s}}$ denotes the vectorization of a symmetric second-order tensor into its independent components, so as to preserve the norm, $\|\mathrm{vec}_\mathrm{s}(\bm A)\|_2 = \|\bm A\|_F$, where $\|\cdot\|_F$ is the Frobenius norm, defined as $(\sum_{\beta}\sum_{\alpha}A_{\alpha\beta}^2)^{1/2}$.
$\mathcal{R}(\bm{\xi})$ denotes the regularization term, which controls the trade-off between goodness of fit and model complexity.
This formulation is a linear regression problem to infer $\bm{\xi}$, where the design matrix has the size of $6N$ rows and $M$ columns.
The design matrix and the target vector are normalized by their norm before performing the regression.

The values of $\bm\xi$ are obtained by the sequentially thresholded ridge (STRidge) algorithm~\cite{Brunton2016-ie} with a ridge weight $h_\mathrm{r}$ and a threshold $h_\mathrm{t}$. 
We first solve the ridge regression problem with a penalty term $\mathcal{R}(\bm{\xi})=h_\mathrm{r}^2\|\bm \xi\|_2^2$.
Coefficients satisfying $|\xi_m|<h_\mathrm{t}$ are then set to zero, and the ridge regression is applied using only the remaining active terms.
This thresholding and refitting procedure is iterated until the nonzero terms in $\bm \xi$ no longer change. 
The resulting $\bm{\xi}$ yields the inferred model.
The hyperparameter set ${h}=\{h_\mathrm{t},h_\mathrm{r}\}$ is selected using a simulation error, which evaluates the time-integrated stress trajectory of the inferred model, as defined in Sec.~\ref{sec:numerical-setup}.

\subsection{Numerical setup and model selection}
\label{sec:numerical-setup}

We construct reduced closures from data generated by two extra-stress models: (I) the original Giesekus model with $\xi_{\mathrm{G}}=0.3$ and (II) the original Larson model with $\xi_{\mathrm{L}}=0.3$.
Their projected closures introduced in Eqs.~\eqref{eq:projected-giesekus} and \eqref{eq:projected-larson} are used as baseline deviatoric-stress models for comparison.

The numerical procedure consists of four steps: (i) generating stress-response data from a prescribed constitutive model, (ii) inferring reduced closures over a grid of STRidge hyperparameters, (iii) selecting the model using simulation errors 
over the training trajectories, and (iv) testing the selected model on deformation protocols not used for training.

(i) Generating the dataset.
We use three deformation types for training: steady shear ($\dot{\gamma}\in\{1,10,100\}$), oscillatory shear ($\dot{\gamma}(t)=\gamma_0\omega\cos(\omega t)$ with $\gamma_0=1$ and $\omega\in\{1,3,10\}$), and steady uniaxial elongation ($\dot{\varepsilon}\in\{1,10,100\}$).
For each deformation protocol, the stress dynamics is simulated from $t=0$ with $\bm{\tau}=\bm{0}$ to $t=10$ with the sampling interval $\Delta t=0.01$, yielding $S_r=1001$ samples per trajectory, and a total of $N=9009$ samples over all $R=9$ trajectories.
The observable variables are $\{\sigma_\mathrm{s},N_1,N_2\}$ for shear and $\sigma_\mathrm{e}$ for uniaxial elongation.
The time derivative $\dot{\bm{s}}$ is obtained directly from the governing dynamical equations, rather than by numerical differentiation of the sampled stress data.

(ii) Performing the regression.
From the generated samples, we reconstruct $\bm{s}$ from the observable stress variables using Eqs.~\eqref{eq:dev-stress-from-shear} and \eqref{eq:dev-stress-from-elongation}, and evaluate the value of $\dev\bm{F}(\bm s,\bm D)$ by using Eq.~\eqref{eq:reduced-closure}.
The invariant-library candidates are chosen as $\phi_{ij}\in\{1,I_2,I_3,I_4,I_5,I_6,I_7,I_8,I_9\}$ for each tensor $\dev\bm{T}_i(\bm s,\bm D)$ with $i=2,\ldots,9$, giving the library size $M=72$.
The identically zero quantities $I_1(\bm s,\bm D)=\tr\bm{s}=0$ and $\dev\bm{T}_1(\bm s,\bm D)=\dev\textbf{I}=0$ are omitted from the library.
Note that higher-order-invariants-included libraries generated inaccurate preliminary predictions due to numerical difficulty, caused by larger magnitudes and non-orthogonalities.
The resulting design matrix contains approximately $10^6$ elements.

(iii) Selecting the model.
We use the STRidge method with the hyperparameter set $h=\{h_\mathrm{t},h_\mathrm{r}\}$.
The trial values are $h_\mathrm{t}\in\{10^{-7},10^{-6},10^{-5},10^{-4},10^{-3},10^{-2},10^{-1}\}$ and $h_\mathrm{r}\in\{10^{-14},10^{-12},10^{-10},10^{-8},10^{-6},10^{-4},10^{-2}\}$, resulting in 49 candidate models.
For each hyperparameter set, the inferred model is evaluated by the normalized simulation error over the training trajectories:
\begin{align}
    E(h)&=\frac{1}{R}\sum_{r=1}^{R}E_r(h),\\
    E_r(h)&=\frac{
    \bigl\{\sum_{l=0}^{S_r-1}\bigl\|\bm{Y}^{(r)}_{\mathrm{pred}}(t_l;h)-\bm{Y}^{(r)}_{\mathrm{true}}(t_l)\bigr\|_2^2 \bigr\}^{1/2}}{\bigl\{\sum_{l=0}^{S_r-1}\bigl\|\bm{Y}^{(r)}_{\mathrm{true}}(t_l)\bigr\|_2^2\bigr\}^{1/2}}.
\end{align}
The hyperparameter set $h^\ast$ was selected to favor sparsity among models with $E(h)$ within 1\% of the minimum.
$\bm{Y}_{\mathrm{true}}^{(r)}$ denotes the reference observable stress trajectory, and $\bm{Y}_{\mathrm{pred}}^{(r)}$ denotes the trajectory obtained by time-integrating the inferred reduced model.
For shear trajectories, $\bm{Y}=[\sigma_\mathrm{s},N_1,N_2]^{\mathsf{T}}$, whereas for uniaxial elongation, $\bm{Y}=\sigma_\mathrm{e}$.

(iv) Testing the selected model.
After selecting the hyperparameters, the final model is assessed on deformation protocols not used for training.
We employ two deformation modes: planar elongation, with $\dot{\varepsilon}_\mathrm{p}=\kappa_{xx}=-\kappa_{yy}$ and $\sigma_\mathrm{p}=s_{xx}-s_{yy}$, and mixed shear and uniaxial elongational deformation, with nonzero $\dot{\gamma}(=\kappa_{xy})$ and $\dot{\varepsilon}(=\kappa_{xx})$.
For planar elongation, we impose a steady deformation with $\dot{\varepsilon}_\mathrm{p}=1$ followed by cessation at $t=5$, and oscillatory deformations $\dot{\varepsilon}_\mathrm{p}(t)=\varepsilon_{p,0}\omega\cos(\omega t)$ with $\varepsilon_{p,0}=1$ and $\omega=1$ and $5$.
For mixed deformation, we impose a steady protocol with $\dot{\gamma}=1$ and $\dot{\varepsilon}=0.3$ followed by cessation at $t=5$, and oscillatory protocols with $(\gamma_0,\varepsilon_0,\omega)=(1,0.3,1)$, $(10, 3, 5)$.
Furthermore, steady shear and uniaxial elongational viscosities are evaluated over $\dot{\gamma},\dot{\varepsilon}\in[10^{-3},10^{3}]$.

\section{Results and Discussion}
\label{sec:resultsanddiscussion}

\subsection{Steady viscosity of the projected closures}
\label{sec:analysis-on-reduced-formula}

\begin{figure*}[htbp]
    \centering
    \includegraphics[width=0.85\linewidth]{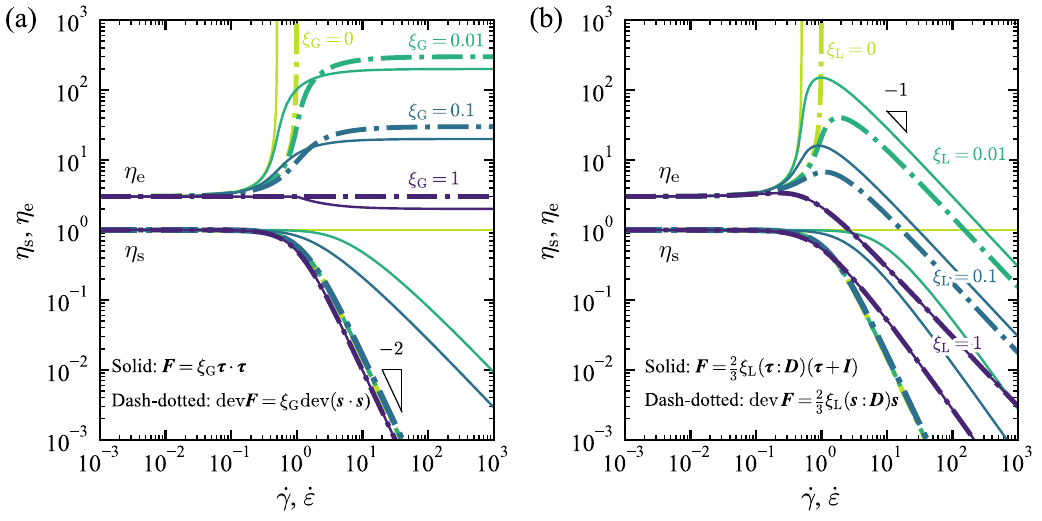}
    \caption{
Steady rheological functions of the original extra-stress models and the projected deviatoric-stress closures for (a) Giesekus and (b) Larson models.
The steady shear viscosity $\eta_\mathrm{s}=\sigma_\mathrm{s}/\dot{\gamma}$ and uniaxial elongational viscosity $\eta_\mathrm{e}=\sigma_\mathrm{e}/\dot{\varepsilon}$ are plotted as functions of the deformation rate.
Dash-dotted curves denote the projected deviatoric-stress closures, whereas solid curves denote the corresponding original extra-stress models.
The parameter values are $\xi_{\mathrm{G}},\xi_{\mathrm{L}}\in\{0,0.01,0.1,1\}$.
}
\label{fig:reduced-steady-comparison}
\end{figure*}

We first examined the steady rheological functions of the projected deviatoric-stress closures.
Figure~\ref{fig:reduced-steady-comparison} compares these projected closures with the corresponding original extra-stress models.
The projected Giesekus and Larson closures are reduced to the $\dev\bm{F}=\bm{0}$ case when $\xi_\mathrm{G}=0$ or $\xi_\mathrm{L}=0$, giving the shear and uniaxial elongational viscosities $\eta_\mathrm{s}=\sigma_\mathrm{s}/\dot\gamma$ and $\eta_\mathrm{e}=\sigma_\mathrm{e}/\dot\varepsilon$:
\begin{equation}
    \eta_\mathrm{s}=\frac{1}{1+\frac{2}{3}\dot\gamma^2},\qquad
    \eta_\mathrm{e}=\frac{3}{1-\dot\varepsilon}.
\end{equation}
The projected closure with $\dev\bm{F}=\bm{0}$ already gives a nonlinear steady shear viscosity although no additional nonlinear term is introduced in $\dev\bm{F}$.
Thus, shear thinning in this projected closure originates from the traceless upper-convected time derivative term in Eq.~\eqref{eq:reduced-closure}, not from a model-specific nonlinear stress contribution.

For finite $\xi_\mathrm{G}$ or $\xi_\mathrm{L}$, the nonlinear terms modify the steady rheological functions in different ways.
In uniaxial elongation, the projected closure with $\dev\bm{F}=\bm{0}$ exhibits a finite-rate divergence at $\dot\varepsilon=1$.
The projected Giesekus and Larson closures remove this divergence, but their high-rate behavior is different.
The projected Giesekus closure gives an extensional saturation, $\eta_{\mathrm{e}}\to 3/\xi_{\mathrm{G}}$, whereas the projected Larson closure gives elongational thinning at large $\dot{\varepsilon}$.

Figure~\ref{fig:reduced-steady-comparison} demonstrates that the projected closures do not reproduce the steady rheological functions of the original models.
This difference was expected because the projected closures are closed equations in the deviatoric-stress state space after eliminating the trace degree of freedom.
For the parameter range shown in Fig.~\ref{fig:reduced-steady-comparison}, the shear response is less sensitive to the choice of the projected nonlinear term than the uniaxial elongational response.
The qualitative shear-thinning behavior is already present in the $\dev\bm{F}=\bm{0}$ projected closure, whereas the nonlinear terms in the Giesekus and Larson models mainly control the elongational response.
These results characterize the steady rheological behavior induced by the projected deviatoric-stress dynamics, which is used as a baseline for data-driven closure inference.

\subsection{Inferred closures}
\label{sec:inferred-model-results}

\begin{figure*}[htbp]
    \centering
    \includegraphics[width=0.9\linewidth]{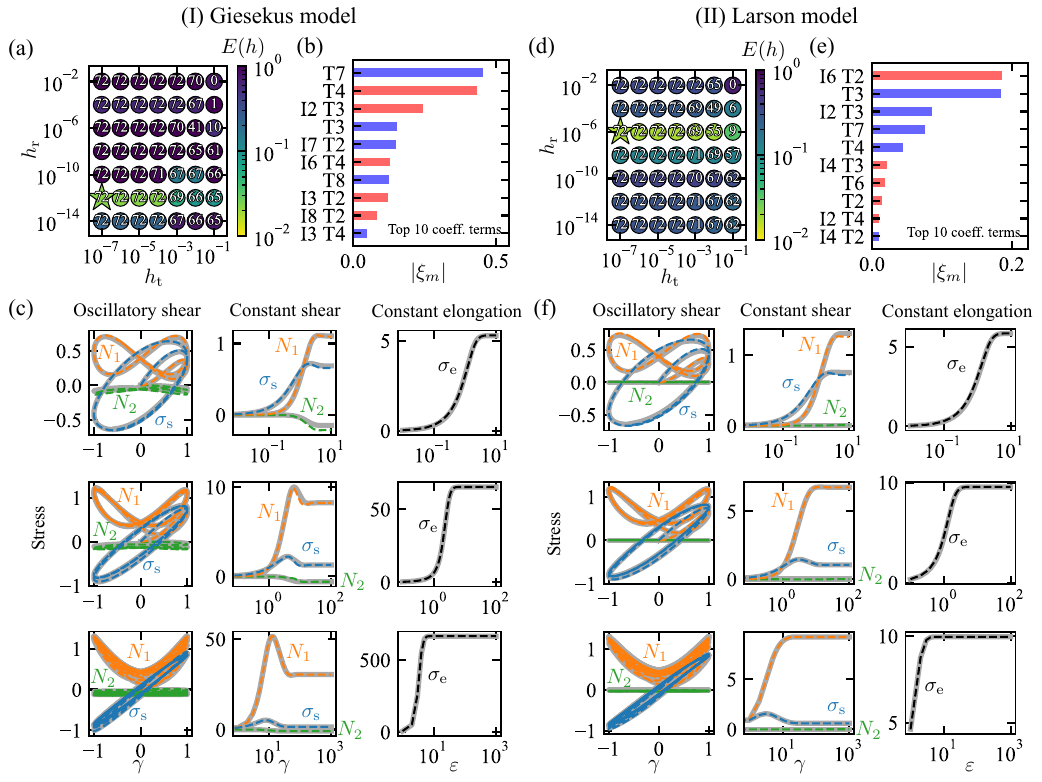}
\caption{
Simulation-error-based inference of deviatoric-stress closures from the original Giesekus and Larson models.
The left and right blocks correspond to data generated from the Giesekus and Larson models with $\xi_\mathrm{G}=\xi_\mathrm{L}=0.3$, respectively.
(a,d) Normalized simulation error $E({h})$ over the STRidge hyperparameters, namely the threshold $h_{\mathrm{t}}$ and ridge weight $h_{\mathrm{r}}$.
The stars denote the selected hyperparameter sets.
The numbers indicate the number of active terms remaining after thresholding, out of the 72 candidate terms.
(b,e) Ten largest-magnitude coefficients $\xi_m$ of the selected closures among the active candidate functions.
Red and blue bars denote positive and negative coefficients, respectively.
(c,f) Reconstruction of the training transients under oscillatory shear, constant shear, and constant uniaxial elongation.
Solid curves denote the original extra-stress models, whereas dashed curves denote the inferred deviatoric-stress closures.
}
\label{fig:inferred-closures}
\end{figure*}

We inferred deviatoric-stress closures from stress-response data generated by the original Giesekus and Larson models.
Figure~\ref{fig:inferred-closures} summarizes the inference results.
Figures~\ref{fig:inferred-closures}(a,d) show the dependence of $E(h)$ on the STRidge hyperparameters.
Although the algorithm prioritizes parsimonious expressions, the closures selected here are not necessarily the most reduced ones, because the present selection criterion prioritizes accuracy in the predicted stress response over the smallest possible support or a smaller regression residual.

Figures~\ref{fig:inferred-closures}(b,e) show the ten largest-magnitude coefficients of the selected closures.
The inferred coefficients are distributed over multiple invariant-tensor candidate functions, suggesting that the reduced closure is consistent with an effective approximation of the eliminated trace memory.
Because several candidate functions can be strongly correlated over the restricted training protocols, the individual coefficients are not structurally identifiable.
For data generated from the original Giesekus model, the projected term $\dev\bm{T}_4=\dev(\bm s\cdot\bm s)$ is included as a candidate, whereas for data generated from the original Larson model, the projected term is $I_6\dev\bm{T}_2=\tr(\bm s\cdot\bm D)\bm s$ (cf. Eqs. \eqref{eq:projected-giesekus} and \eqref{eq:projected-larson}).

Also in Figs.~\ref{fig:inferred-closures}(b,e), some large coefficients are associated with terms proportional to $\dev\bm{T}_2=\bm{s}$ and $\dev\bm{T}_3=\bm{D}$.
These terms have the same tensorial forms as the linear relaxation and driving terms in Eq.~\eqref{eq:reduced-closure}, and can act as effective renormalizations of the linear response in the reduced state space.
Thus, the inferred coefficient vector should be interpreted as an effective closure for observable deviatoric-stress dynamics under the trained deformations, rather than as a unique decomposition of linear/nonlinear physical mechanisms.

Figures~\ref{fig:inferred-closures}(c,f) compare the transient stress responses obtained from the original extra-stress models and the inferred deviatoric-stress closures under the deformation conditions used for the training data.
For both datasets, the inferred closures reproduced the stress responses under oscillatory shear, constant shear, and constant uniaxial elongation.
In shear, the shear stress $\sigma_{\mathrm{s}}$ and normal stress differences $N_1$ and $N_2$ were reconstructed, while in uniaxial elongation the elongational stress $\sigma_{\mathrm{e}}$ was reconstructed up to $\dot\gamma,\dot\varepsilon=100/\lambda$.
These results indicate that the observable deviatoric-stress dynamics generated by the original extra-stress models can be represented by closed reduced dynamics over the training deformation protocols.

\subsection{Extrapolation and limitations}
\label{sec:inferred-model-test}

\begin{figure*}[htbp]
    \centering
    \includegraphics[width=0.9\linewidth]{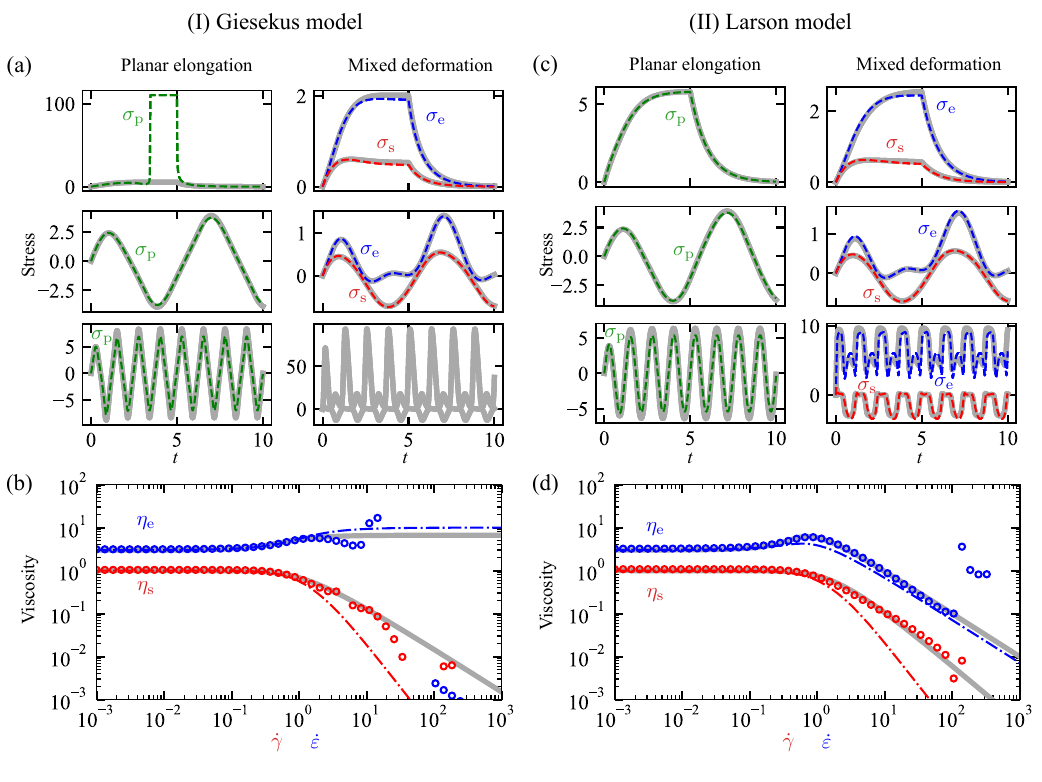}
\caption{
Transient stress responses and steady rheological functions of the inferred deviatoric-stress closures under test deformations.
The left and right blocks correspond to closures inferred from data generated by the original Giesekus and Larson models, respectively.
The details of the applied strain rate histories are written in Sec.~\ref{sec:numerical-setup}.
(a,c) Test transients under planar elongation and mixed shear and uniaxial elongational deformation.
For each deformation mode, cessation and oscillatory protocols are shown.
These deformation protocols were not used for training and spanned moderate to strong deformation rates.
Solid curves denote the original extra-stress models, whereas dashed curves denote the inferred deviatoric-stress closures.
(b,d) Steady shear viscosity $\eta_{\mathrm{s}}=\sigma_{\mathrm{s}}/\dot{\gamma}$ and uniaxial elongational viscosity $\eta_{\mathrm{e}}=\sigma_{\mathrm{e}}/\dot{\varepsilon}$ obtained from the selected closures.
Thick solid curves denote the original extra-stress models, dash-dotted curves denote the projected reduced closures, and circles denote steady-state values obtained from simulations of the inferred closures.
In the bottom-right panel of (I-a), the predicted stress responses diverge numerically, and the corresponding curves are not shown.
The inferred closures reproduce the original extra-stress models over part of the deformation-rate range connected to the training data, whereas deviations and divergent steady responses appear at larger deformation rates.
}
\label{fig:test-rheology}
\end{figure*}

We evaluated the selected closures on deformation protocols that were not used for training. 
Figure~\ref{fig:test-rheology} summarizes the validation using untrained transient flows and steady rheological functions. 
Figures~\ref{fig:test-rheology}(a,c) show transient responses under planar elongation and mixed shear/uniaxial elongations, including cessation and oscillatory protocols spanning deformation rates from order $\lambda^{-1}$ to substantially higher rates. 
Figures~\ref{fig:test-rheology}(b,d) compare the steady shear viscosity $\eta_\mathrm{s}=\sigma_\mathrm{s}/\dot{\gamma}$ and uniaxial elongational viscosity $\eta_\mathrm{e}=\sigma_\mathrm{e}/\dot{\epsilon}$ from the original, inferred, and projected models. 
This comparison clarifies the role of inferred terms by symbolic regression beyond the simply projected closures examined in Fig.~\ref{fig:reduced-steady-comparison}. 
Although the projected reduced closures mainly modify the steady elongational responses and leave the steady shear responses nearly unchanged, the inferred models can follow the original extra-stress models over the deformation-rate range.

The transient deformation tests under planar elongation and mixed deformation indicate that the inferred closures retain transferability beyond the training deformation modes. 
Although trained only on shear and uniaxial elongational responses, the closures capture the main stress transients under planar elongation and mixed shear and uniaxial elongational deformation at moderate deformation rates, particularly in Fig.~\ref{fig:test-rheology}(c). 
This transferability is nevertheless limited when the test protocol involves large extensional stress growth. 
For the data generated from the original Giesekus model, the training response contains large elongational stresses under steady uniaxial elongation, as shown in Fig.~\ref{fig:inferred-closures}(c). 
Although the inferred closure reproduces this training trajectory, it shows deviations under mixed shear and uniaxial elongational deformation in Fig.~\ref{fig:test-rheology}(a).
By contrast, the data from the Larson model involve smaller elongational stress magnitudes $\sigma_\mathrm{e}\sim\mathcal{O}(1)$, and the corresponding transient responses remain more stable, compared to the Giesekus model $\sigma_\mathrm{e}\sim\mathcal{O}(\dot\varepsilon)$. 
This difference suggests that the robustness of the inferred model depends on the large stress scale in the training data, the resulting contributions of higher-order candidate terms, and the simulation-error-based model-selection procedure.
The steady rheological functions lead to the same conclusion. 
As shown in Figs.~\ref{fig:test-rheology}(b,d), the inferred closures reproduce the original extra-stress models over part of the deformation-rate range connected to the training data, whereas deviations and divergent steady responses appear at larger deformation rates. 

This limitation was expected because the present inference constrains finite-time trajectories over prescribed training protocols, but does not impose the correct high-rate asymptotic behavior or bounded steady responses outside the trained range. 
Thus, the inferred equations should be interpreted as effective reduced closures for observable deviatoric-stress dynamics within the range covered by the training data. 
Improving the strong-deformation behavior would require additional training data emphasizing extensional responses, structural constraints on the closure form, a library designed to encode the expected asymptotic behavior, and careful normalization of simulation-error-based model selection to avoid overemphasizing large-magnitude stress responses.

\section{Conclusion}

We proposed a viscoelastic constitutive formulation written in terms of deviatoric stress, to describe experimentally accessible rheological data.
Standard rheological measurements provide only selected stress components and do not determine the isotropic part of the extra stress tensor. 
In the proposed formulation, the deviatoric stress is reconstructed from observable quantities, including the shear stress, normal stress differences, and elongational stress, and its dynamics is represented by a closure inferred by a regression protocol.
The proposed reduction procedure preserves the tensorial structure, material objectivity, and linear Maxwell-type response while avoiding the unobserved trace degree of freedom.

The inferred closures reproduced the training responses and generalized to untrained deformation protocols over deformation rates comparable to the inverse relaxation time.
However, the steady rheological functions deviated from the reference models, and in some cases diverged, at larger deformation rates outside the training regime.
Although the present proof of concept uses noiseless model-generated data and exact stress derivatives, application to experimental data will additionally require treatment of measurement noise, temporal differentiation, and incomplete availability of the second normal stress difference.
These limitations indicate the need for stable model selection, improved numerical implementation, and physically informed restriction of candidate invariants and tensor bases.
A broader description of linear viscoelasticity will additionally require a multimode extension or an explicit memory-kernel representation.

Despite these limitations, the present study provides a constitutive modeling framework based solely on deviatoric stress.
The proposed framework offers a practical route to inferring objective tensorial dynamics from experimentally accessible rheological data.

\begin{acknowledgments}
This work was supported by the Japan Society for the Promotion of Science (JSPS) KAKENHI under Grant No.~25K23525 awarded to S.M. and 26K01281 to T.S.
\end{acknowledgments}

\appendix

\section{Latent-state estimation for partially observed extra-stress tensor dynamics}
\label{sec:latent-estimation}

In many rheological measurements, not all components of $\bm{\tau}$ are directly observable.
Thus, we outline an extension of the symbolic regression framework to the partially observed setting, where the constitutive coefficients and the unobserved stress components are inferred simultaneously.~\cite{Brunton2019-gp,Lu2022-co}

Let $\bm{X}$ denote a state vector that collects the independent stress components relevant to the given deformation modes, and let $\bm{U}$ denote the corresponding reduced input.
The full tensors $\bm{\tau}$ and $\bm{\kappa}$ are reconstructed from these variables as
\begin{align}
    \mathrm{vec}_\mathrm{s}(\bm \tau) = \bm P_{\bm\tau}\bm X,\\
    \mathrm{vec}(\bm \kappa) = \bm P_{\bm\kappa}\bm U.
\end{align}
Here, $\mathrm{vec}_\mathrm{s}$ has been defined as Eq.~\eqref{eq:vec-sym}, and $\mathrm{vec}(\bm{B})$ is written as
\begin{equation}
    \mathrm{vec}(\bm B)=[B_{xx},B_{xy},B_{xz},B_{yx},B_{yy},B_{yz},B_{zx},B_{zy},B_{zz}]^\mathsf{T},
\end{equation}
where $\bm P_{\bm\tau}$ and $\bm P_{\bm\kappa}$ are fixed linear maps.
We collect the experimentally accessible quantities in an observation vector $\bm{Y}$, which typically has fewer degrees of freedom than $\bm{X}$.
The observation and reconstruction are written as
\begin{align}
    &\bm Y=\bm P\bm X,\\
    &\bm X=\bm P^{\dagger}\bm Y+  \bm N \bm Z.
\end{align}
Here, $\bm{P}$ is the observation matrix and $\bm{P}^{\dagger}$ is a right pseudoinverse of $\bm{P}$.
The matrix $\bm{N}$ spans the null space of $\bm{P}$, satisfying $\bm{P}\bm{N}=\bm{0}$ and $\bm{N}^{\mathsf{T}}\bm{N}=\bm{I}$, and $\bm{Z}$ is the latent-state vector parameterizing the unobserved degrees of freedom.

Instead of estimating $\bm{Z}$ independently at each time point, we parameterize each latent trajectory $\bm{Z}^{(r)}(t)$ using basis-spline coefficients for $r=1,\ldots,R$:
\begin{equation}
    {\bm Z}^{(r)}(t)=\sum_{l=1}^L {\bm c_\ell^{(r)}} {\bm b}_\ell(t),
\end{equation} 
where $L$ is the total number of expansion functions, ${\bm c}^{(r)}_\ell$ and ${\bm b}_\ell$ are $\ell$-th coefficient vectors and spline functions, respectively.
For compactness, we define ${\bm c}^{(r)}=[(\bm c_1^{(r)})^\mathsf{T}\ \ldots\ ({\bm c}_L^{(r)})^\mathsf{T}]^\mathsf{T}$ and $\bm c=[(\bm c^{(1)})^\mathsf{T}\ \ldots\ (\bm c^{(R)})^\mathsf{T}]^\mathsf{T}$.

In this partially observed formulation, both the residual vector $\bm{f}_{\tau}^{(r,p)}$ and the design matrix block $\bm{\theta}_{\tau}^{(r,p)}$ depend on the latent coefficients $\bm{c}^{(r)}$.
We denote the $p$-th sample of the $r$-th trajectory by $(r,p)$, where $p=1,\ldots,S_r$, and the total number of samples is $N=\sum_{r=1}^{R}S_r$.
The constitutive coefficients $\bm{\xi}$ and latent coefficients $\bm{c}$ are estimated simultaneously by minimizing the objective function with the residual and regularization terms:
\begin{align}
    \mathcal{J}(\bm \xi,\bm c)=&
\frac{1}{N}\sum_{r=1}^R\sum_{p=1}^{S_r}\|\bm{f}_{\bm\tau}^{(r,p)}(\bm c^{(r)})-\bm{\theta}_{\bm\tau}^{(r,p)}(\bm c^{(r)})\bm{\xi}\|_2^2 \nonumber\\
&+ \mathcal{R}(\bm{\xi}),
\end{align}
where 
\begin{align}
\bm{f}_{\bm\tau}^{(r,p)}
&=
\mathrm{vec}_{\mathrm{s}}(\bm{F}^{(r,p)}), \\
\bm{\theta}_{\bm\tau}^{(r,p)}
&=
\bigl[
\mathrm{vec}_{\mathrm{s}}(\bm{\Psi}_1^{(r,p)})\,
\ldots\,
\mathrm{vec}_{\mathrm{s}}(\bm{\Psi}_M^{(r,p)})
\bigr].    
\end{align}
Here, $\bm{F}^{(r,p)}$ is evaluated as the residual of the extra-stress equation for the reconstructed $\bm{\tau}^{(r,p)}$ and $\bm{\kappa}^{(r,p)}$.

Algorithmically, one can alternately update $\bm{\xi}$ with $\bm{c}$ fixed and update $\bm{c}$ with $\bm{\xi}$ fixed.
The former step is the same symbolic-regression problem as in the fully observed case, whereas the latter requires a nonlinear optimization over the latent coefficients $\bm{c}$.
Such a solver can be implemented by combining a linear regression solver for $\bm{\xi}$ with a gradient-based optimizer for $\bm{c}$ using automatic differentiation.

For example, under simple shear deformation in the $x$--$y$ plane with $\kappa_{xy}$, the nonzero components of $\bm{\tau}$ are $\tau_{xy}$ and the three diagonal components.
The observable quantities are the shear stress $\sigma_{\mathrm{s}}=\tau_{xy}$, the first normal stress difference $N_1=\tau_{xx}-\tau_{yy}$, and the second normal stress difference $N_2=\tau_{yy}-\tau_{zz}$.
For this $(\sigma_{\mathrm{s}},N_1,N_2)$-observed shear case, the vectors and linear maps are written as
\begin{align}
    &\bm X=\begin{bmatrix}\tau_{xx} & \tau_{yy}& \tau_{zz}& \tau_{xy}\end{bmatrix}^\mathsf{T},\\
    &U=\kappa_{xy},\\
    &\bm P_{\bm \tau}=
    \begin{bmatrix}
        1&0&0&0\\
        0&1&0&0\\
        0&0&1&0\\
        0&0&0&\sqrt{2}\\
        0&0&0&0\\
        0&0&0&0
    \end{bmatrix},\\
    &\bm P_{\bm \kappa}=\begin{bmatrix}0& 1& 0& 0& 0& 0& 0& 0& 0\end{bmatrix}^\mathsf{T}.
\end{align}
Here, $U$ is a scalar input.
The observation and latent variables are
\begin{align}
    &\bm Y=\begin{bmatrix}\sigma_\mathrm{s} & N_1 & N_2\end{bmatrix}^\mathsf{T},\\
    &Z =\tr(\bm\tau)/\sqrt{3},\\
    &\bm P=
    \begin{bmatrix}
        0&0&0&1\\
        1&-1&0&0\\
        0&1&-1&0
    \end{bmatrix},\ 
    \bm P^\dagger=
    \begin{bmatrix}
    0&{2}/{3}&{1}/{3}\\
    0&-{1}/{3}&{1}/{3}\\
    0&-{1}/{3}&-{2}/{3}\\
    1&0&0
    \end{bmatrix},\\
    &\bm N=\begin{bmatrix}1/\sqrt{3}\ 1/\sqrt{3}\ 1/\sqrt{3}\ 0\end{bmatrix}^\mathsf{T}.
\end{align}
This example shows that the trace mode is the latent degree of freedom left by the standard shear observables.

Although this latent-stress reconstruction provides a formal route to extra-stress constitutive inference under partial observation, this approach is highly ill-conditioned in practice.
The latent trace trajectory can absorb nonlinear stress contributions, leading to non-unique or unstable symbolic representations.
We do not use this formulation in the main analysis of this paper.
Instead, we formulate the problem directly in terms of the observable deviatoric-stress dynamics.

Note that this formulation differs from the partial-information setting of Shanbhag and Erlebacher~\cite{Shanbhag2024-po}. 
Their method is formulated for oscillatory shear and fits the Fourier coefficients of the periodic shear-stress response predicted by a tensor-basis constitutive model; unmeasured normal stresses are generated internally during the forward solution. 
In contrast, the formulation considered here treats the unobserved trace of the extra stress as a latent trajectory and can, in principle, be applied to different deformation modes through the corresponding observation and reconstruction maps. 
This deformation-mode generality comes at the cost of a more nonunique and ill-conditioned joint estimation problem for the latent trajectory and constitutive coefficients. 
The same identifiability issue can arise in alternative formulations that integrate the latent $z$-dynamics while simultaneously infferring $\bm F$, because the inferred trace contribution and latent trajectory compensate for each other. 

\section{Exact trace elimination in the UCM model}
\label{sec:trace-elimination-UCM}

The UCM model provides a simple example showing that the deviatoric reduction is not a direct projection of the nonlinear term in the extra-stress equation. For the extra-stress UCM model,
\begin{equation}
\overset{\triangledown}{\bm\tau}+\bm\tau=2\bm D ,
\end{equation}
we decompose the stress tensor as $\bm\tau=\bm s+z\textbf{I}/3$, where $z=\mathrm{tr}\bm\tau$ and $\bm s=\dev\bm\tau$. 
The trace and deviatoric parts are then written as
\begin{align}
&\dot z + z = 2\bm D:\bm s,\\
&\overset{\triangledown_0}{\bm s} + \bm s - \frac{2}{3}z\bm D = 2\bm D .
\end{align}
The trace equation gives
\begin{equation}
z(t) = z(0)e^{-t}+ 2\int^t_0 e^{-(t-t^\prime)}\bm s(t^\prime):\bm D(t^\prime)dt^\prime.
\end{equation}
For the initial condition, $z(0)=0$, substitution into the deviatoric equation yields the trace-eliminated form:
\begin{equation}
\dev\bm F =-\frac{2}{3}z\bm D=-\frac{4}{3}\bm D\int^t_0 e^{-(t-t^\prime)}\bm s(t^\prime):\bm D(t^\prime)dt^\prime .
\end{equation}
This result shows that eliminating the trace mode generates a memory contribution in the deviatoric dynamics, even when the original extra-stress model has no nonlinear term. 
The present \(\bm s\)-closure should be regarded as a Markovian approximation of the original extra-stress formula.

\onecolumngrid

\bibliographystyle{apsrev4-2}
\bibliography{ref}

\begin{thebibliography}{47}%
\makeatletter
\providecommand \@ifxundefined [1]{%
 \@ifx{#1\undefined}
}%
\providecommand \@ifnum [1]{%
 \ifnum #1\expandafter \@firstoftwo
 \else \expandafter \@secondoftwo
 \fi
}%
\providecommand \@ifx [1]{%
 \ifx #1\expandafter \@firstoftwo
 \else \expandafter \@secondoftwo
 \fi
}%
\providecommand \natexlab [1]{#1}%
\providecommand \enquote  [1]{``#1''}%
\providecommand \bibnamefont  [1]{#1}%
\providecommand \bibfnamefont [1]{#1}%
\providecommand \citenamefont [1]{#1}%
\providecommand \href@noop [0]{\@secondoftwo}%
\providecommand \href [0]{\begingroup \@sanitize@url \@href}%
\providecommand \@href[1]{\@@startlink{#1}\@@href}%
\providecommand \@@href[1]{\endgroup#1\@@endlink}%
\providecommand \@sanitize@url [0]{\catcode `\\12\catcode `\$12\catcode
  `\&12\catcode `\#12\catcode `\^12\catcode `\_12\catcode `\%12\relax}%
\providecommand \@@startlink[1]{}%
\providecommand \@@endlink[0]{}%
\providecommand \url  [0]{\begingroup\@sanitize@url \@url }%
\providecommand \@url [1]{\endgroup\@href {#1}{\urlprefix }}%
\providecommand \urlprefix  [0]{URL }%
\providecommand \Eprint [0]{\href }%
\providecommand \doibase [0]{https://doi.org/}%
\providecommand \selectlanguage [0]{\@gobble}%
\providecommand \bibinfo  [0]{\@secondoftwo}%
\providecommand \bibfield  [0]{\@secondoftwo}%
\providecommand \translation [1]{[#1]}%
\providecommand \BibitemOpen [0]{}%
\providecommand \bibitemStop [0]{}%
\providecommand \bibitemNoStop [0]{.\EOS\space}%
\providecommand \EOS [0]{\spacefactor3000\relax}%
\providecommand \BibitemShut  [1]{\csname bibitem#1\endcsname}%
\let\auto@bib@innerbib\@empty
\bibitem [{\citenamefont {Doi}\ and\ \citenamefont
  {Edwards}(1988)}]{Doi1988-yf}%
  \BibitemOpen
  \bibfield  {author} {\bibinfo {author} {\bibfnamefont {M.}~\bibnamefont
  {Doi}}\ and\ \bibinfo {author} {\bibfnamefont {S.~F.}\ \bibnamefont
  {Edwards}},\ }\href@noop {} {\emph {\bibinfo {title} {The theory of polymer
  dynamics}}}\ (\bibinfo  {publisher} {Clarendon Press},\ \bibinfo {year}
  {1988})\BibitemShut {NoStop}%
\bibitem [{\citenamefont {Larson}(1998)}]{Larson1998-ia}%
  \BibitemOpen
  \bibfield  {author} {\bibinfo {author} {\bibfnamefont {R.~G.}\ \bibnamefont
  {Larson}},\ }\href@noop {} {\emph {\bibinfo {title} {The structure and
  rheology of complex fluids}}}\ (\bibinfo  {publisher} {Oxford University
  Press},\ \bibinfo {year} {1998})\BibitemShut {NoStop}%
\bibitem [{\citenamefont {Oldroyd}(1958)}]{Oldroyd1958-jh}%
  \BibitemOpen
  \bibfield  {author} {\bibinfo {author} {\bibfnamefont {J.~G.}\ \bibnamefont
  {Oldroyd}},\ }\href {https://doi.org/10.1098/rspa.1958.0083} {\bibfield
  {journal} {\bibinfo  {journal} {Proc. R. Soc. Lond.}\ }\textbf {\bibinfo
  {volume} {245}},\ \bibinfo {pages} {278} (\bibinfo {year}
  {1958})}\BibitemShut {NoStop}%
\bibitem [{\citenamefont {Giesekus}(1982)}]{Giesekus1982-ov}%
  \BibitemOpen
  \bibfield  {author} {\bibinfo {author} {\bibfnamefont {H.}~\bibnamefont
  {Giesekus}},\ }\href {https://doi.org/10.1016/0377-0257(82)85016-7}
  {\bibfield  {journal} {\bibinfo  {journal} {J. Nonnewton. Fluid Mech.}\
  }\textbf {\bibinfo {volume} {11}},\ \bibinfo {pages} {69} (\bibinfo {year}
  {1982})}\BibitemShut {NoStop}%
\bibitem [{\citenamefont {Thien}\ and\ \citenamefont
  {Tanner}(1977)}]{Thien1977-gs}%
  \BibitemOpen
  \bibfield  {author} {\bibinfo {author} {\bibfnamefont {N.~P.}\ \bibnamefont
  {Thien}}\ and\ \bibinfo {author} {\bibfnamefont {R.~I.}\ \bibnamefont
  {Tanner}},\ }\href {https://doi.org/10.1016/0377-0257(77)80021-9} {\bibfield
  {journal} {\bibinfo  {journal} {J. Nonnewton. Fluid Mech.}\ }\textbf
  {\bibinfo {volume} {2}},\ \bibinfo {pages} {353} (\bibinfo {year}
  {1977})}\BibitemShut {NoStop}%
\bibitem [{\citenamefont {Larson}(1984)}]{Larson1984-hw}%
  \BibitemOpen
  \bibfield  {author} {\bibinfo {author} {\bibfnamefont {R.~G.}\ \bibnamefont
  {Larson}},\ }\href {https://doi.org/10.1122/1.549761} {\bibfield  {journal}
  {\bibinfo  {journal} {J. Rheol.}\ }\textbf {\bibinfo {volume} {28}},\
  \bibinfo {pages} {545} (\bibinfo {year} {1984})}\BibitemShut {NoStop}%
\bibitem [{\citenamefont {Peterlin}(1966)}]{Peterlin1966-xc}%
  \BibitemOpen
  \bibfield  {author} {\bibinfo {author} {\bibfnamefont {A.}~\bibnamefont
  {Peterlin}},\ }\href {https://doi.org/10.1002/pol.1966.110040411} {\bibfield
  {journal} {\bibinfo  {journal} {J. Polym. Sci. B}\ }\textbf {\bibinfo
  {volume} {4}},\ \bibinfo {pages} {287} (\bibinfo {year} {1966})}\BibitemShut
  {NoStop}%
\bibitem [{\citenamefont {Hyun}\ \emph {et~al.}(2011)\citenamefont {Hyun},
  \citenamefont {Wilhelm}, \citenamefont {Klein}, \citenamefont {Cho},
  \citenamefont {Nam}, \citenamefont {Ahn}, \citenamefont {Lee}, \citenamefont
  {Ewoldt},\ and\ \citenamefont {McKinley}}]{Hyun2011-xf}%
  \BibitemOpen
  \bibfield  {author} {\bibinfo {author} {\bibfnamefont {K.}~\bibnamefont
  {Hyun}}, \bibinfo {author} {\bibfnamefont {M.}~\bibnamefont {Wilhelm}},
  \bibinfo {author} {\bibfnamefont {C.~O.}\ \bibnamefont {Klein}}, \bibinfo
  {author} {\bibfnamefont {K.~S.}\ \bibnamefont {Cho}}, \bibinfo {author}
  {\bibfnamefont {J.~G.}\ \bibnamefont {Nam}}, \bibinfo {author} {\bibfnamefont
  {K.~H.}\ \bibnamefont {Ahn}}, \bibinfo {author} {\bibfnamefont {S.~J.}\
  \bibnamefont {Lee}}, \bibinfo {author} {\bibfnamefont {R.~H.}\ \bibnamefont
  {Ewoldt}},\ and\ \bibinfo {author} {\bibfnamefont {G.~H.}\ \bibnamefont
  {McKinley}},\ }\href {https://doi.org/10.1016/j.progpolymsci.2011.02.002}
  {\bibfield  {journal} {\bibinfo  {journal} {Prog. Polym. Sci.}\ }\textbf
  {\bibinfo {volume} {36}},\ \bibinfo {pages} {1697} (\bibinfo {year}
  {2011})}\BibitemShut {NoStop}%
\bibitem [{\citenamefont {Bharadwaj}\ and\ \citenamefont
  {Ewoldt}(2015)}]{Bharadwaj2015-ov}%
  \BibitemOpen
  \bibfield  {author} {\bibinfo {author} {\bibfnamefont {N.~A.}\ \bibnamefont
  {Bharadwaj}}\ and\ \bibinfo {author} {\bibfnamefont {R.~H.}\ \bibnamefont
  {Ewoldt}},\ }\href {https://doi.org/10.1122/1.4903346} {\bibfield  {journal}
  {\bibinfo  {journal} {J. Rheol.}\ }\textbf {\bibinfo {volume} {59}},\
  \bibinfo {pages} {557} (\bibinfo {year} {2015})}\BibitemShut {NoStop}%
\bibitem [{\citenamefont {Saengow}\ \emph {et~al.}(2015)\citenamefont
  {Saengow}, \citenamefont {Giacomin},\ and\ \citenamefont
  {Kolitawong}}]{Saengow2015-np}%
  \BibitemOpen
  \bibfield  {author} {\bibinfo {author} {\bibfnamefont {C.}~\bibnamefont
  {Saengow}}, \bibinfo {author} {\bibfnamefont {A.~J.}\ \bibnamefont
  {Giacomin}},\ and\ \bibinfo {author} {\bibfnamefont {C.}~\bibnamefont
  {Kolitawong}},\ }\href {https://doi.org/10.1002/mats.201400104} {\bibfield
  {journal} {\bibinfo  {journal} {Macromol. Theory Simul.}\ }\textbf {\bibinfo
  {volume} {24}},\ \bibinfo {pages} {352} (\bibinfo {year} {2015})}\BibitemShut
  {NoStop}%
\bibitem [{\citenamefont {Saengow}\ and\ \citenamefont
  {Giacomin}(2018)}]{Saengow2018-hx}%
  \BibitemOpen
  \bibfield  {author} {\bibinfo {author} {\bibfnamefont {C.}~\bibnamefont
  {Saengow}}\ and\ \bibinfo {author} {\bibfnamefont {A.~J.}\ \bibnamefont
  {Giacomin}},\ }\href {https://doi.org/10.1063/1.5023586} {\bibfield
  {journal} {\bibinfo  {journal} {Phys. Fluids}\ }\textbf {\bibinfo {volume}
  {30}},\ \bibinfo {pages} {030703} (\bibinfo {year} {2018})}\BibitemShut
  {NoStop}%
\bibitem [{\citenamefont {Saengow}\ \emph {et~al.}(2019)\citenamefont
  {Saengow}, \citenamefont {Giacomin}, \citenamefont {Grizzuti},\ and\
  \citenamefont {Pasquino}}]{Saengow2019-uw}%
  \BibitemOpen
  \bibfield  {author} {\bibinfo {author} {\bibfnamefont {C.}~\bibnamefont
  {Saengow}}, \bibinfo {author} {\bibfnamefont {A.~J.}\ \bibnamefont
  {Giacomin}}, \bibinfo {author} {\bibfnamefont {N.}~\bibnamefont {Grizzuti}},\
  and\ \bibinfo {author} {\bibfnamefont {R.}~\bibnamefont {Pasquino}},\ }\href
  {https://doi.org/10.1063/1.5091493} {\bibfield  {journal} {\bibinfo
  {journal} {Phys. Fluids}\ }\textbf {\bibinfo {volume} {31}},\ \bibinfo
  {pages} {063101} (\bibinfo {year} {2019})}\BibitemShut {NoStop}%
\bibitem [{\citenamefont {Song}\ \emph {et~al.}(2020)\citenamefont {Song},
  \citenamefont {Kong}, \citenamefont {Kim},\ and\ \citenamefont
  {Hyun}}]{Song2020-ln}%
  \BibitemOpen
  \bibfield  {author} {\bibinfo {author} {\bibfnamefont {H.~Y.}\ \bibnamefont
  {Song}}, \bibinfo {author} {\bibfnamefont {H.~J.}\ \bibnamefont {Kong}},
  \bibinfo {author} {\bibfnamefont {S.~Y.}\ \bibnamefont {Kim}},\ and\ \bibinfo
  {author} {\bibfnamefont {K.}~\bibnamefont {Hyun}},\ }\href
  {https://doi.org/10.1122/1.5139685} {\bibfield  {journal} {\bibinfo
  {journal} {J. Rheol.}\ }\textbf {\bibinfo {volume} {64}},\ \bibinfo {pages}
  {673} (\bibinfo {year} {2020})}\BibitemShut {NoStop}%
\bibitem [{\citenamefont {Saadat}\ \emph {et~al.}(2022)\citenamefont {Saadat},
  \citenamefont {Mahmoudabadbozchelou},\ and\ \citenamefont
  {Jamali}}]{Saadat2022-lk}%
  \BibitemOpen
  \bibfield  {author} {\bibinfo {author} {\bibfnamefont {M.}~\bibnamefont
  {Saadat}}, \bibinfo {author} {\bibfnamefont {M.}~\bibnamefont
  {Mahmoudabadbozchelou}},\ and\ \bibinfo {author} {\bibfnamefont
  {S.}~\bibnamefont {Jamali}},\ }\href
  {https://doi.org/10.1007/s00397-022-01357-w} {\bibfield  {journal} {\bibinfo
  {journal} {Rheol. Acta}\ }\textbf {\bibinfo {volume} {61}},\ \bibinfo {pages}
  {721} (\bibinfo {year} {2022})}\BibitemShut {NoStop}%
\bibitem [{\citenamefont {Dabiri}\ \emph {et~al.}(2023)\citenamefont {Dabiri},
  \citenamefont {Saadat}, \citenamefont {Mangal},\ and\ \citenamefont
  {Jamali}}]{Dabiri2023-st}%
  \BibitemOpen
  \bibfield  {author} {\bibinfo {author} {\bibfnamefont {D.}~\bibnamefont
  {Dabiri}}, \bibinfo {author} {\bibfnamefont {M.}~\bibnamefont {Saadat}},
  \bibinfo {author} {\bibfnamefont {D.}~\bibnamefont {Mangal}},\ and\ \bibinfo
  {author} {\bibfnamefont {S.}~\bibnamefont {Jamali}},\ }\href
  {https://doi.org/10.1007/s00397-023-01408-w} {\bibfield  {journal} {\bibinfo
  {journal} {Rheol. Acta}\ }\textbf {\bibinfo {volume} {62}},\ \bibinfo {pages}
  {557} (\bibinfo {year} {2023})}\BibitemShut {NoStop}%
\bibitem [{\citenamefont {John}\ \emph {et~al.}(2024)\citenamefont {John},
  \citenamefont {Mowbray}, \citenamefont {Alalwyat}, \citenamefont
  {Vousvoukis}, \citenamefont {Martin}, \citenamefont {Kowalski},\ and\
  \citenamefont {Fonte}}]{John2024-bu}%
  \BibitemOpen
  \bibfield  {author} {\bibinfo {author} {\bibfnamefont {T.~P.}\ \bibnamefont
  {John}}, \bibinfo {author} {\bibfnamefont {M.}~\bibnamefont {Mowbray}},
  \bibinfo {author} {\bibfnamefont {A.}~\bibnamefont {Alalwyat}}, \bibinfo
  {author} {\bibfnamefont {M.}~\bibnamefont {Vousvoukis}}, \bibinfo {author}
  {\bibfnamefont {P.}~\bibnamefont {Martin}}, \bibinfo {author} {\bibfnamefont
  {A.}~\bibnamefont {Kowalski}},\ and\ \bibinfo {author} {\bibfnamefont
  {C.~P.}\ \bibnamefont {Fonte}},\ }\href
  {https://doi.org/10.1016/j.ces.2024.120075} {\bibfield  {journal} {\bibinfo
  {journal} {Chem. Eng. Sci.}\ }\textbf {\bibinfo {volume} {294}},\ \bibinfo
  {pages} {120075} (\bibinfo {year} {2024})}\BibitemShut {NoStop}%
\bibitem [{\citenamefont {Saadat}\ \emph {et~al.}(2023)\citenamefont {Saadat},
  \citenamefont {Mangal},\ and\ \citenamefont {Jamali}}]{Saadat2023-pg}%
  \BibitemOpen
  \bibfield  {author} {\bibinfo {author} {\bibfnamefont {M.}~\bibnamefont
  {Saadat}}, \bibinfo {author} {\bibfnamefont {D.}~\bibnamefont {Mangal}},\
  and\ \bibinfo {author} {\bibfnamefont {S.}~\bibnamefont {Jamali}},\ }\href
  {https://doi.org/10.1039/d3dd00036b} {\bibfield  {journal} {\bibinfo
  {journal} {Digit. Discov.}\ }\textbf {\bibinfo {volume} {2}},\ \bibinfo
  {pages} {915} (\bibinfo {year} {2023})}\BibitemShut {NoStop}%
\bibitem [{\citenamefont {Brunton}\ and\ \citenamefont
  {Kutz}(2019)}]{Brunton2019-gp}%
  \BibitemOpen
  \bibfield  {author} {\bibinfo {author} {\bibfnamefont {S.~L.}\ \bibnamefont
  {Brunton}}\ and\ \bibinfo {author} {\bibfnamefont {J.~N.}\ \bibnamefont
  {Kutz}},\ }\href {https://doi.org/10.1017/9781108380690} {\emph {\bibinfo
  {title} {Data-driven science and engineering}}}\ (\bibinfo  {publisher}
  {Cambridge University Press},\ \bibinfo {year} {2019})\BibitemShut {NoStop}%
\bibitem [{\citenamefont {Vinuesa}\ and\ \citenamefont
  {Brunton}(2022)}]{Vinuesa2022-bm}%
  \BibitemOpen
  \bibfield  {author} {\bibinfo {author} {\bibfnamefont {R.}~\bibnamefont
  {Vinuesa}}\ and\ \bibinfo {author} {\bibfnamefont {S.~L.}\ \bibnamefont
  {Brunton}},\ }\href {https://doi.org/10.1038/s43588-022-00264-7} {\bibfield
  {journal} {\bibinfo  {journal} {Nat. Comput. Sci.}\ }\textbf {\bibinfo
  {volume} {2}},\ \bibinfo {pages} {358} (\bibinfo {year} {2022})}\BibitemShut
  {NoStop}%
\bibitem [{\citenamefont {Mangal}\ \emph
  {et~al.}(2025{\natexlab{a}})\citenamefont {Mangal}, \citenamefont {Jha},
  \citenamefont {Dabiri},\ and\ \citenamefont {Jamali}}]{Mangal2024-gi}%
  \BibitemOpen
  \bibfield  {author} {\bibinfo {author} {\bibfnamefont {D.}~\bibnamefont
  {Mangal}}, \bibinfo {author} {\bibfnamefont {A.}~\bibnamefont {Jha}},
  \bibinfo {author} {\bibfnamefont {D.}~\bibnamefont {Dabiri}},\ and\ \bibinfo
  {author} {\bibfnamefont {S.}~\bibnamefont {Jamali}},\ }\href
  {https://doi.org/10.1016/j.cocis.2024.101873} {\bibfield  {journal} {\bibinfo
   {journal} {Curr. Opin. Colloid Interface Sci.}\ }\textbf {\bibinfo {volume}
  {75}},\ \bibinfo {pages} {101873} (\bibinfo {year}
  {2025}{\natexlab{a}})}\BibitemShut {NoStop}%
\bibitem [{\citenamefont {Mahmoudabadbozchelou}\ \emph
  {et~al.}(2022)\citenamefont {Mahmoudabadbozchelou}, \citenamefont {Kamani},
  \citenamefont {Rogers},\ and\ \citenamefont
  {Jamali}}]{Mahmoudabadbozchelou2022-er}%
  \BibitemOpen
  \bibfield  {author} {\bibinfo {author} {\bibfnamefont {M.}~\bibnamefont
  {Mahmoudabadbozchelou}}, \bibinfo {author} {\bibfnamefont {K.~M.}\
  \bibnamefont {Kamani}}, \bibinfo {author} {\bibfnamefont {S.~A.}\
  \bibnamefont {Rogers}},\ and\ \bibinfo {author} {\bibfnamefont
  {S.}~\bibnamefont {Jamali}},\ }\href
  {https://doi.org/10.1073/pnas.2202234119} {\bibfield  {journal} {\bibinfo
  {journal} {Proc. Natl. Acad. Sci. U. S. A.}\ }\textbf {\bibinfo {volume}
  {119}},\ \bibinfo {pages} {e2202234119} (\bibinfo {year} {2022})}\BibitemShut
  {NoStop}%
\bibitem [{\citenamefont {Mahmoudabadbozchelou}\ \emph
  {et~al.}(2024)\citenamefont {Mahmoudabadbozchelou}, \citenamefont {Kamani},
  \citenamefont {Rogers},\ and\ \citenamefont
  {Jamali}}]{Mahmoudabadbozchelou2024-zi}%
  \BibitemOpen
  \bibfield  {author} {\bibinfo {author} {\bibfnamefont {M.}~\bibnamefont
  {Mahmoudabadbozchelou}}, \bibinfo {author} {\bibfnamefont {K.~M.}\
  \bibnamefont {Kamani}}, \bibinfo {author} {\bibfnamefont {S.~A.}\
  \bibnamefont {Rogers}},\ and\ \bibinfo {author} {\bibfnamefont
  {S.}~\bibnamefont {Jamali}},\ }\href
  {https://doi.org/10.1073/pnas.2313658121} {\bibfield  {journal} {\bibinfo
  {journal} {Proc. Natl. Acad. Sci. U. S. A.}\ }\textbf {\bibinfo {volume}
  {121}},\ \bibinfo {pages} {e2313658121} (\bibinfo {year} {2024})}\BibitemShut
  {NoStop}%
\bibitem [{\citenamefont {Seryo}\ \emph {et~al.}(2020)\citenamefont {Seryo},
  \citenamefont {Sato}, \citenamefont {Molina},\ and\ \citenamefont
  {Taniguchi}}]{Seryo2020-ib}%
  \BibitemOpen
  \bibfield  {author} {\bibinfo {author} {\bibfnamefont {N.}~\bibnamefont
  {Seryo}}, \bibinfo {author} {\bibfnamefont {T.}~\bibnamefont {Sato}},
  \bibinfo {author} {\bibfnamefont {J.~J.}\ \bibnamefont {Molina}},\ and\
  \bibinfo {author} {\bibfnamefont {T.}~\bibnamefont {Taniguchi}},\ }\href
  {https://doi.org/10.1103/physrevresearch.2.033107} {\bibfield  {journal}
  {\bibinfo  {journal} {Phys. Rev. Res.}\ }\textbf {\bibinfo {volume} {2}},\
  \bibinfo {pages} {033107} (\bibinfo {year} {2020})}\BibitemShut {NoStop}%
\bibitem [{\citenamefont {Miyamoto}\ \emph {et~al.}(2023)\citenamefont
  {Miyamoto}, \citenamefont {Molina},\ and\ \citenamefont
  {Taniguchi}}]{Miyamoto2023-ot}%
  \BibitemOpen
  \bibfield  {author} {\bibinfo {author} {\bibfnamefont {S.}~\bibnamefont
  {Miyamoto}}, \bibinfo {author} {\bibfnamefont {J.~J.}\ \bibnamefont
  {Molina}},\ and\ \bibinfo {author} {\bibfnamefont {T.}~\bibnamefont
  {Taniguchi}},\ }\href {https://doi.org/10.1063/5.0156272} {\bibfield
  {journal} {\bibinfo  {journal} {Phys. Fluids}\ }\textbf {\bibinfo {volume}
  {35}},\ \bibinfo {pages} {063113} (\bibinfo {year} {2023})}\BibitemShut
  {NoStop}%
\bibitem [{\citenamefont {Zhao}\ \emph {et~al.}(2021)\citenamefont {Zhao},
  \citenamefont {Li}, \citenamefont {Wang}, \citenamefont {Caswell},
  \citenamefont {Ouyang},\ and\ \citenamefont {Karniadakis}}]{Zhao2021-kl}%
  \BibitemOpen
  \bibfield  {author} {\bibinfo {author} {\bibfnamefont {L.}~\bibnamefont
  {Zhao}}, \bibinfo {author} {\bibfnamefont {Z.}~\bibnamefont {Li}}, \bibinfo
  {author} {\bibfnamefont {Z.}~\bibnamefont {Wang}}, \bibinfo {author}
  {\bibfnamefont {B.}~\bibnamefont {Caswell}}, \bibinfo {author} {\bibfnamefont
  {J.}~\bibnamefont {Ouyang}},\ and\ \bibinfo {author} {\bibfnamefont {G.~E.}\
  \bibnamefont {Karniadakis}},\ }\href
  {https://doi.org/10.1016/j.jcp.2020.110069} {\bibfield  {journal} {\bibinfo
  {journal} {J. Comput. Phys.}\ }\textbf {\bibinfo {volume} {427}},\ \bibinfo
  {pages} {110069} (\bibinfo {year} {2021})}\BibitemShut {NoStop}%
\bibitem [{\citenamefont {Jin}\ \emph {et~al.}(2023)\citenamefont {Jin},
  \citenamefont {Yoon}, \citenamefont {Park},\ and\ \citenamefont
  {Ahn}}]{Jin2023-ki}%
  \BibitemOpen
  \bibfield  {author} {\bibinfo {author} {\bibfnamefont {H.}~\bibnamefont
  {Jin}}, \bibinfo {author} {\bibfnamefont {S.}~\bibnamefont {Yoon}}, \bibinfo
  {author} {\bibfnamefont {F.~C.}\ \bibnamefont {Park}},\ and\ \bibinfo
  {author} {\bibfnamefont {K.~H.}\ \bibnamefont {Ahn}},\ }\href
  {https://doi.org/10.1007/s00397-023-01405-z} {\bibfield  {journal} {\bibinfo
  {journal} {Rheol. Acta}\ }\textbf {\bibinfo {volume} {62}},\ \bibinfo {pages}
  {569} (\bibinfo {year} {2023})}\BibitemShut {NoStop}%
\bibitem [{\citenamefont {Lennon}\ \emph {et~al.}(2023)\citenamefont {Lennon},
  \citenamefont {McKinley},\ and\ \citenamefont {Swan}}]{Lennon2023-tt}%
  \BibitemOpen
  \bibfield  {author} {\bibinfo {author} {\bibfnamefont {K.~R.}\ \bibnamefont
  {Lennon}}, \bibinfo {author} {\bibfnamefont {G.~H.}\ \bibnamefont
  {McKinley}},\ and\ \bibinfo {author} {\bibfnamefont {J.~W.}\ \bibnamefont
  {Swan}},\ }\href {https://doi.org/10.1073/pnas.2304669120} {\bibfield
  {journal} {\bibinfo  {journal} {Proc. Natl. Acad. Sci. U. S. A.}\ }\textbf
  {\bibinfo {volume} {120}},\ \bibinfo {pages} {e2304669120} (\bibinfo {year}
  {2023})}\BibitemShut {NoStop}%
\bibitem [{\citenamefont {Shanbhag}\ and\ \citenamefont
  {Erlebacher}(2024)}]{Shanbhag2024-po}%
  \BibitemOpen
  \bibfield  {author} {\bibinfo {author} {\bibfnamefont {S.}~\bibnamefont
  {Shanbhag}}\ and\ \bibinfo {author} {\bibfnamefont {G.}~\bibnamefont
  {Erlebacher}},\ }\href {https://doi.org/10.1063/5.0233607} {\bibfield
  {journal} {\bibinfo  {journal} {Phys. Fluids}\ }\textbf {\bibinfo {volume}
  {36}},\ \bibinfo {pages} {103117} (\bibinfo {year} {2024})}\BibitemShut
  {NoStop}%
\bibitem [{\citenamefont {Sato}\ and\ \citenamefont
  {Miyamoto}(2025)}]{Sato2025-yx}%
  \BibitemOpen
  \bibfield  {author} {\bibinfo {author} {\bibfnamefont {T.}~\bibnamefont
  {Sato}}\ and\ \bibinfo {author} {\bibfnamefont {S.}~\bibnamefont
  {Miyamoto}},\ }\href {https://doi.org/10.1007/s00397-025-01491-1} {\bibfield
  {journal} {\bibinfo  {journal} {Rheol. Acta}\ }\textbf {\bibinfo {volume}
  {64}},\ \bibinfo {pages} {443} (\bibinfo {year} {2025})}\BibitemShut
  {NoStop}%
\bibitem [{\citenamefont {Sato}\ \emph {et~al.}(2025)\citenamefont {Sato},
  \citenamefont {Miyamoto},\ and\ \citenamefont {Kato}}]{Sato2025-eh}%
  \BibitemOpen
  \bibfield  {author} {\bibinfo {author} {\bibfnamefont {T.}~\bibnamefont
  {Sato}}, \bibinfo {author} {\bibfnamefont {S.}~\bibnamefont {Miyamoto}},\
  and\ \bibinfo {author} {\bibfnamefont {S.}~\bibnamefont {Kato}},\ }\href
  {https://doi.org/10.1122/8.0000872} {\bibfield  {journal} {\bibinfo
  {journal} {J. Rheol.}\ }\textbf {\bibinfo {volume} {69}},\ \bibinfo {pages}
  {15} (\bibinfo {year} {2025})}\BibitemShut {NoStop}%
\bibitem [{\citenamefont {Rodrigues}\ \emph {et~al.}(2025)\citenamefont
  {Rodrigues}, \citenamefont {Thompson}, \citenamefont {Oliveira},\ and\
  \citenamefont {Ausas}}]{Rodrigues2025-ke}%
  \BibitemOpen
  \bibfield  {author} {\bibinfo {author} {\bibfnamefont {E.~C.}\ \bibnamefont
  {Rodrigues}}, \bibinfo {author} {\bibfnamefont {R.~L.}\ \bibnamefont
  {Thompson}}, \bibinfo {author} {\bibfnamefont {D.~A.~B.}\ \bibnamefont
  {Oliveira}},\ and\ \bibinfo {author} {\bibfnamefont {R.~F.}\ \bibnamefont
  {Ausas}},\ }\href {https://doi.org/10.1016/j.engappai.2025.111788} {\bibfield
   {journal} {\bibinfo  {journal} {Eng. Appl. Artif. Intell.}\ }\textbf
  {\bibinfo {volume} {160}},\ \bibinfo {pages} {111788} (\bibinfo {year}
  {2025})}\BibitemShut {NoStop}%
\bibitem [{\citenamefont {Mangal}\ \emph
  {et~al.}(2025{\natexlab{b}})\citenamefont {Mangal}, \citenamefont {Saadat},\
  and\ \citenamefont {Jamali}}]{Mangal2025-fo}%
  \BibitemOpen
  \bibfield  {author} {\bibinfo {author} {\bibfnamefont {D.}~\bibnamefont
  {Mangal}}, \bibinfo {author} {\bibfnamefont {M.}~\bibnamefont {Saadat}},\
  and\ \bibinfo {author} {\bibfnamefont {S.}~\bibnamefont {Jamali}},\ }\href
  {https://doi.org/10.1122/8.0000908} {\bibfield  {journal} {\bibinfo
  {journal} {J. Rheol.}\ }\textbf {\bibinfo {volume} {69}},\ \bibinfo {pages}
  {55} (\bibinfo {year} {2025}{\natexlab{b}})}\BibitemShut {NoStop}%
\bibitem [{\citenamefont {Sato}\ \emph {et~al.}(2026)\citenamefont {Sato},
  \citenamefont {Miyamoto},\ and\ \citenamefont {Kato}}]{Sato2026-fk}%
  \BibitemOpen
  \bibfield  {author} {\bibinfo {author} {\bibfnamefont {T.}~\bibnamefont
  {Sato}}, \bibinfo {author} {\bibfnamefont {S.}~\bibnamefont {Miyamoto}},\
  and\ \bibinfo {author} {\bibfnamefont {S.}~\bibnamefont {Kato}},\ }\href
  {https://doi.org/10.1678/rheology.54.119} {\bibfield  {journal} {\bibinfo
  {journal} {Nihon Reoroji Gakkaishi}\ }\textbf {\bibinfo {volume} {54}},\
  \bibinfo {pages} {119} (\bibinfo {year} {2026})}\BibitemShut {NoStop}%
\bibitem [{\citenamefont {Fang}\ \emph {et~al.}(2022)\citenamefont {Fang},
  \citenamefont {Ge}, \citenamefont {Zhang}, \citenamefont {E},\ and\
  \citenamefont {Lei}}]{Fang2024-vx}%
  \BibitemOpen
  \bibfield  {author} {\bibinfo {author} {\bibfnamefont {L.}~\bibnamefont
  {Fang}}, \bibinfo {author} {\bibfnamefont {P.}~\bibnamefont {Ge}}, \bibinfo
  {author} {\bibfnamefont {L.}~\bibnamefont {Zhang}}, \bibinfo {author}
  {\bibfnamefont {W.}~\bibnamefont {E}},\ and\ \bibinfo {author} {\bibfnamefont
  {H.}~\bibnamefont {Lei}},\ }\href {https://doi.org/10.4208/jml.220115}
  {\bibfield  {journal} {\bibinfo  {journal} {J. Mach. Learn.}\ }\textbf
  {\bibinfo {volume} {1}},\ \bibinfo {pages} {114} (\bibinfo {year}
  {2022})}\BibitemShut {NoStop}%
\bibitem [{\citenamefont {Dong}\ \emph {et~al.}(2026)\citenamefont {Dong},
  \citenamefont {Simavilla}, \citenamefont {Ouyang}, \citenamefont {Wang},\
  and\ \citenamefont {Ellero}}]{Dong2026-ic}%
  \BibitemOpen
  \bibfield  {author} {\bibinfo {author} {\bibfnamefont {X.}~\bibnamefont
  {Dong}}, \bibinfo {author} {\bibfnamefont {D.~N.}\ \bibnamefont {Simavilla}},
  \bibinfo {author} {\bibfnamefont {J.}~\bibnamefont {Ouyang}}, \bibinfo
  {author} {\bibfnamefont {X.}~\bibnamefont {Wang}},\ and\ \bibinfo {author}
  {\bibfnamefont {M.}~\bibnamefont {Ellero}},\ }\href
  {https://doi.org/10.1016/j.jcp.2026.114837} {\bibfield  {journal} {\bibinfo
  {journal} {J. Comput. Phys.}\ }\textbf {\bibinfo {volume} {557}},\ \bibinfo
  {pages} {114837} (\bibinfo {year} {2026})}\BibitemShut {NoStop}%
\bibitem [{\citenamefont {Park}\ \emph {et~al.}(2026)\citenamefont {Park},
  \citenamefont {Yoon}, \citenamefont {Jin}, \citenamefont {Park},\ and\
  \citenamefont {Myung}}]{Park2026-zw}%
  \BibitemOpen
  \bibfield  {author} {\bibinfo {author} {\bibfnamefont {N.}~\bibnamefont
  {Park}}, \bibinfo {author} {\bibfnamefont {S.}~\bibnamefont {Yoon}}, \bibinfo
  {author} {\bibfnamefont {H.}~\bibnamefont {Jin}}, \bibinfo {author}
  {\bibfnamefont {J.~D.}\ \bibnamefont {Park}},\ and\ \bibinfo {author}
  {\bibfnamefont {J.~S.}\ \bibnamefont {Myung}},\ }\href
  {https://doi.org/10.1007/s00397-026-01585-4} {\bibfield  {journal} {\bibinfo
  {journal} {Rheol. Acta}\ }\textbf {\bibinfo {volume} {65}},\ \bibinfo {pages}
  {785} (\bibinfo {year} {2026})}\BibitemShut {NoStop}%
\bibitem [{\citenamefont {Schweizer}\ and\ \citenamefont
  {Schmidheiny}(2013)}]{Schweizer2013-om}%
  \BibitemOpen
  \bibfield  {author} {\bibinfo {author} {\bibfnamefont {T.}~\bibnamefont
  {Schweizer}}\ and\ \bibinfo {author} {\bibfnamefont {W.}~\bibnamefont
  {Schmidheiny}},\ }\href {https://doi.org/10.1122/1.4797458} {\bibfield
  {journal} {\bibinfo  {journal} {J. Rheol.}\ }\textbf {\bibinfo {volume}
  {57}},\ \bibinfo {pages} {841} (\bibinfo {year} {2013})}\BibitemShut
  {NoStop}%
\bibitem [{\citenamefont {Costanzo}\ \emph {et~al.}(2018)\citenamefont
  {Costanzo}, \citenamefont {Ianniruberto}, \citenamefont {Marrucci},\ and\
  \citenamefont {Vlassopoulos}}]{Costanzo2018-sc}%
  \BibitemOpen
  \bibfield  {author} {\bibinfo {author} {\bibfnamefont {S.}~\bibnamefont
  {Costanzo}}, \bibinfo {author} {\bibfnamefont {G.}~\bibnamefont
  {Ianniruberto}}, \bibinfo {author} {\bibfnamefont {G.}~\bibnamefont
  {Marrucci}},\ and\ \bibinfo {author} {\bibfnamefont {D.}~\bibnamefont
  {Vlassopoulos}},\ }\href {https://doi.org/10.1007/s00397-018-1080-1}
  {\bibfield  {journal} {\bibinfo  {journal} {Rheol. Acta}\ }\textbf {\bibinfo
  {volume} {57}},\ \bibinfo {pages} {363} (\bibinfo {year} {2018})}\BibitemShut
  {NoStop}%
\bibitem [{\citenamefont {Costanzo}\ \emph {et~al.}(2024)\citenamefont
  {Costanzo}, \citenamefont {Parisi}, \citenamefont {Schweizer},\ and\
  \citenamefont {Vlassopoulos}}]{Costanzo2024-cb}%
  \BibitemOpen
  \bibfield  {author} {\bibinfo {author} {\bibfnamefont {S.}~\bibnamefont
  {Costanzo}}, \bibinfo {author} {\bibfnamefont {D.}~\bibnamefont {Parisi}},
  \bibinfo {author} {\bibfnamefont {T.}~\bibnamefont {Schweizer}},\ and\
  \bibinfo {author} {\bibfnamefont {D.}~\bibnamefont {Vlassopoulos}},\ }\href
  {https://doi.org/10.1122/8.0000897} {\bibfield  {journal} {\bibinfo
  {journal} {J. Rheol.}\ }\textbf {\bibinfo {volume} {68}},\ \bibinfo {pages}
  {1013} (\bibinfo {year} {2024})}\BibitemShut {NoStop}%
\bibitem [{\citenamefont {McKinley}\ and\ \citenamefont
  {Tripathi}(2000)}]{McKinley2000-ow}%
  \BibitemOpen
  \bibfield  {author} {\bibinfo {author} {\bibfnamefont {G.~H.}\ \bibnamefont
  {McKinley}}\ and\ \bibinfo {author} {\bibfnamefont {A.}~\bibnamefont
  {Tripathi}},\ }\href {https://doi.org/10.1122/1.551105} {\bibfield  {journal}
  {\bibinfo  {journal} {J. Rheol.}\ }\textbf {\bibinfo {volume} {44}},\
  \bibinfo {pages} {653} (\bibinfo {year} {2000})}\BibitemShut {NoStop}%
\bibitem [{\citenamefont {Haward}\ \emph {et~al.}(2012)\citenamefont {Haward},
  \citenamefont {Oliveira}, \citenamefont {Alves},\ and\ \citenamefont
  {McKinley}}]{Haward2012-px}%
  \BibitemOpen
  \bibfield  {author} {\bibinfo {author} {\bibfnamefont {S.~J.}\ \bibnamefont
  {Haward}}, \bibinfo {author} {\bibfnamefont {M.~S.~N.}\ \bibnamefont
  {Oliveira}}, \bibinfo {author} {\bibfnamefont {M.~A.}\ \bibnamefont
  {Alves}},\ and\ \bibinfo {author} {\bibfnamefont {G.~H.}\ \bibnamefont
  {McKinley}},\ }\href {https://doi.org/10.1103/PhysRevLett.109.128301}
  {\bibfield  {journal} {\bibinfo  {journal} {Phys. Rev. Lett.}\ }\textbf
  {\bibinfo {volume} {109}},\ \bibinfo {pages} {128301} (\bibinfo {year}
  {2012})}\BibitemShut {NoStop}%
\bibitem [{\citenamefont {Haward}\ \emph
  {et~al.}(2023{\natexlab{a}})\citenamefont {Haward}, \citenamefont {Pimenta},
  \citenamefont {Varchanis}, \citenamefont {Carlson}, \citenamefont
  {Toda-Peters}, \citenamefont {Alves},\ and\ \citenamefont
  {Shen}}]{Haward2023-bd}%
  \BibitemOpen
  \bibfield  {author} {\bibinfo {author} {\bibfnamefont {S.~J.}\ \bibnamefont
  {Haward}}, \bibinfo {author} {\bibfnamefont {F.}~\bibnamefont {Pimenta}},
  \bibinfo {author} {\bibfnamefont {S.}~\bibnamefont {Varchanis}}, \bibinfo
  {author} {\bibfnamefont {D.~W.}\ \bibnamefont {Carlson}}, \bibinfo {author}
  {\bibfnamefont {K.}~\bibnamefont {Toda-Peters}}, \bibinfo {author}
  {\bibfnamefont {M.~A.}\ \bibnamefont {Alves}},\ and\ \bibinfo {author}
  {\bibfnamefont {A.~Q.}\ \bibnamefont {Shen}},\ }\href
  {https://doi.org/10.1122/8.0000659} {\bibfield  {journal} {\bibinfo
  {journal} {J. Rheol.}\ }\textbf {\bibinfo {volume} {67}},\ \bibinfo {pages}
  {995} (\bibinfo {year} {2023}{\natexlab{a}})}\BibitemShut {NoStop}%
\bibitem [{\citenamefont {Haward}\ \emph
  {et~al.}(2023{\natexlab{b}})\citenamefont {Haward}, \citenamefont
  {Varchanis}, \citenamefont {McKinley}, \citenamefont {Alves},\ and\
  \citenamefont {Shen}}]{Haward2023-lg}%
  \BibitemOpen
  \bibfield  {author} {\bibinfo {author} {\bibfnamefont {S.~J.}\ \bibnamefont
  {Haward}}, \bibinfo {author} {\bibfnamefont {S.}~\bibnamefont {Varchanis}},
  \bibinfo {author} {\bibfnamefont {G.~H.}\ \bibnamefont {McKinley}}, \bibinfo
  {author} {\bibfnamefont {M.~A.}\ \bibnamefont {Alves}},\ and\ \bibinfo
  {author} {\bibfnamefont {A.~Q.}\ \bibnamefont {Shen}},\ }\href
  {https://doi.org/10.1122/8.0000660} {\bibfield  {journal} {\bibinfo
  {journal} {J. Rheol.}\ }\textbf {\bibinfo {volume} {67}},\ \bibinfo {pages}
  {1011} (\bibinfo {year} {2023}{\natexlab{b}})}\BibitemShut {NoStop}%
\bibitem [{\citenamefont {Spencer}(2004)}]{Spencer2004-es}%
  \BibitemOpen
  \bibfield  {author} {\bibinfo {author} {\bibfnamefont {A.~J.~M.}\
  \bibnamefont {Spencer}},\ }\href@noop {} {\emph {\bibinfo {title} {Continuum
  Mechanics}}},\ Dover Books on Physics\ (\bibinfo  {publisher} {Dover
  Publications},\ \bibinfo {year} {2004})\BibitemShut {NoStop}%
\bibitem [{\citenamefont {Giacomin}\ and\ \citenamefont
  {Saengow}(2022)}]{Giacomin2022-sy}%
  \BibitemOpen
  \bibfield  {author} {\bibinfo {author} {\bibfnamefont {A.~J.}\ \bibnamefont
  {Giacomin}}\ and\ \bibinfo {author} {\bibfnamefont {C.}~\bibnamefont
  {Saengow}},\ }\href {https://doi.org/10.1016/j.jnnfm.2021.104653} {\bibfield
  {journal} {\bibinfo  {journal} {J. Nonnewton. Fluid Mech.}\ }\textbf
  {\bibinfo {volume} {299}},\ \bibinfo {pages} {104653} (\bibinfo {year}
  {2022})}\BibitemShut {NoStop}%
\bibitem [{\citenamefont {Brunton}\ \emph {et~al.}(2016)\citenamefont
  {Brunton}, \citenamefont {Proctor},\ and\ \citenamefont
  {Kutz}}]{Brunton2016-ie}%
  \BibitemOpen
  \bibfield  {author} {\bibinfo {author} {\bibfnamefont {S.~L.}\ \bibnamefont
  {Brunton}}, \bibinfo {author} {\bibfnamefont {J.~L.}\ \bibnamefont
  {Proctor}},\ and\ \bibinfo {author} {\bibfnamefont {J.~N.}\ \bibnamefont
  {Kutz}},\ }\href {https://doi.org/10.1073/pnas.1517384113} {\bibfield
  {journal} {\bibinfo  {journal} {Proc. Natl. Acad. Sci. U. S. A.}\ }\textbf
  {\bibinfo {volume} {113}},\ \bibinfo {pages} {3932} (\bibinfo {year}
  {2016})}\BibitemShut {NoStop}%
\bibitem [{\citenamefont {Lu}\ \emph {et~al.}(2022)\citenamefont {Lu},
  \citenamefont {Ari\~{n}o Bernad},\ and\ \citenamefont
  {Solja\v{c}i\'{c}}}]{Lu2022-co}%
  \BibitemOpen
  \bibfield  {author} {\bibinfo {author} {\bibfnamefont {P.~Y.}\ \bibnamefont
  {Lu}}, \bibinfo {author} {\bibfnamefont {J.}~\bibnamefont {Ari\~{n}o
  Bernad}},\ and\ \bibinfo {author} {\bibfnamefont {M.}~\bibnamefont
  {Solja\v{c}i\'{c}}},\ }\href {https://doi.org/10.1038/s42005-022-00987-z}
  {\bibfield  {journal} {\bibinfo  {journal} {Commun. Phys.}\ }\textbf
  {\bibinfo {volume} {5}},\ \bibinfo {pages} {206} (\bibinfo {year}
  {2022})}\BibitemShut {NoStop}%
\end{thebibliography}%

\end{document}